\documentstyle[12pt,amssymb,latexsym,amsmath]{article}
\newcommand{\RR}{{\mathbb R}}

\newcommand{\beq}{\begin{equation}}
\newcommand{\eeq}{\end{equation}}
\newcommand{\ba}{\begin{array}}
\newcommand{\ea}{\end{array}}
\begin{document}

\title{Matrix equations of hydrodynamic type as 
lower-dimensional reductions of Self-dual type $S$-integrable systems}

\author{
A.I. Zenchuk\\
Center of Nonlinear Studies of L.D.Landau Institute
for Theoretical Physics  \\
(International Institute of Nonlinear Science)\\
Kosygina 2, Moscow, Russia 119334\\
E-mail: zenchuk@itp.ac.ru
}

\maketitle

\begin{abstract}
We show that  matrix $Q\times Q$ Self-dual type $S$-integrable Partial Differential
 Equations (PDEs) possess a family of  
 lower-dimensional reductions represented by the  matrix 
 $ Q \times n_0 Q$  quasilinear first order PDEs solved in 
 \cite{SZ1} by the method of characteristics. In turn, these PDEs admit two types of available particular solutions: (a)
  explicit solutions and (b) solutions described implicitly by a system  of 
  non-differential equations.  The later 
 solutions, in particular, exhibit the wave profile breaking. 
  Only first type of solutions is available for (1+1)-dimensional nonlinear $S$-integrable  PDEs.   (1+1)-dimensional
 $N$-wave equation, (2+1)- and (3+1)-dimensional Pohlmeyer equations are represented as 
  examples. We also represent a new version of the dressing method which supplies both classical solutions and solutions with wave profile breaking to the above $S$-integrable PDEs. 
\end{abstract}


\section{Introduction}

There are several types of nonlinear equations of Mathematical Physics, which are referred to as  integrable Partial Differential Equations (PDEs). We underline three following 
types of equations:   
\begin{enumerate}
\item
Equations integrable by the Inverse Spectral Transform Method (ISTM), which are also called  soliton or $S$-integrable equations. Korteweg-de-Vries  equation (KdV) has been discovered as the first representative of this class \cite{GGKM}. Among other methods of study of soliton equations we underline so-called dressing method, which was invented in \cite{ZSh1,ZSh2} and  developed in  set of papers, for instance,
 \cite{ZM,BM,Zakharov,Zakharov1,Zakharov2}, see also \cite{AKNS,ZMNP,K}.
 \item
 Nonlinear PDEs linearizable by some direct substitution, or $C$-integrable equations 
 \cite{Calogero,C_int1,C_int2,C_int3,C_int4,C_int5}. Mostly remarkable is Hoph substitution allowing to integrate B\"urgers type nonlinear PDEs.
 \item
 Quasilinear first order  PDEs  integrable by the method of characteristics \cite{Whitham} and by generalized hodograph method \cite{ts2,ts1,dn}. The later method is well applicable to (1+1)-dimensional nonlinear PDEs. Investigation of the higher dimensional systems by this method requires special efforts (see, for instance, the method of hydrodynamic reductions, which will be cited below).
\end{enumerate}

Many other method for study of the nonlinear PDEs have been developed. For instance, the algebraic-geometrical approach is well suited for the construction of {\it local} solutions  \cite{Kr1,Kr2}. However, only  restricted class of global solutions may be 
described in this way. A method for implicit description of solutions manifold is the method of hydrodynamic reductions \cite{Ferapontov1,Ferapontov2,Ferapontov3,Ferapontov4,
Ferapontov5}, where solutions of multidimensional integrable PDEs are described in terms of Riemann invariants of some (1+1)-dimensional first order quasilinear PDEs. The criterion for applicability of such technique to the given nonlinear PDE has been invented. This method allows to implement infinitely many arbitrary functions of {\it single} argument into solution  of the nonlinear PDE satisfying this criterion. 

Our approach to the problem of  construction of  the particular solutions to  the Self-dual type  $S$-integrable PDEs  is most similar to the method of hydrodynamic reductions. However, if it is applicable, we are able to describe the bigger solutions space. Namely, number of arguments in the arbitrary functions implemented into solution has no restriction and is defined by  the particular nonlinear PDE. But  we have no criterion for applicability of this algorithm to the given nonlinear PDE, unlike the method of hydrodynamic reductions. 

We exhibit a  class of solutions with  wave profile breaking  to the  $S$-integrable PDEs selected above. This phenomenon is not surprising, since, simultaneously, these  solutions are solutions of the well established class of first order quasilinear PDEs in lower dimension, which  are known to  possess  solutions spaces implicitly described by systems of non-differential equations \cite{SZ1,ZS2}.  
Remark that solutions with breaking of wave profile  to   dispersion-less Kadomtsev-Pertviashvili equation (dKP) (which is compatibility condition of appropriate pair of vector fields) have been obtained and described in  \cite{SM} using ISTM. 

In the rest of Introduction we, first of all, recall the method of characteristics applicable to a certain type of matrix first order quasilinear PDEs,
 Sec.\ref{Intr:char}.
Then we remind some important properties of the Self-dual type nonlinear $S$-integrable PDEs, Sec.\ref{Intr:Sint}. Basic novalties appearing in this paper  are  underlined in Sec.\ref{Intr:nov}. Finally, we represent the brief contents of  the paper in Sec.\ref{Contents}.  


\subsection{First order quasilinear PDEs integrable by the method of characteristics}
\label{Intr:char}
As it was mentioned above, we associate the  given Self-dual type $S$-integrable nonlinear PDE  with the family of lower dimensional first order matrix PDEs integrable by the method of characteristics. 
Let us describe this family of PDEs. Throughout this paper we will write 
superscripts in parentheses in order to distinguish them from powers. 

It is well known that the scalar first order PDE 
\begin{equation}
\label{equ-u}
u_t+\sum\limits_{k=1}^N u_{x_k}\rho^{(k)}(u)=\rho(u),~~u:\RR^{n+1}\to\RR,
\eeq 
can be solved  by the method of characteristics
\cite{Whitham}. For instance,
 solution $u(x,t)$ of (\ref{equ-u}) for $\rho=0$ is defined
implicitly by the non-differential equation \beq \label{sol-u}
u=f\left(x_1-\rho^{(1)}(u)t,..,x_N-\rho^{(N)}(u)t\right). \eeq
Vector generalizations of equation (\ref{equ-u}), i.e. the systems of
several  coupled first order scalar PDEs,  have been
investigated in set of papers   \cite{ts2,dn,K1,K2}
using generalized hodograph method in most cases.
Recently \cite{SZ1} the matrix generalization of eq.(\ref{equ-u}) has been solved  using algebraic approach:
\begin{eqnarray}
\label{equ-w}
w_t+\sum\limits_{k=1}^N w_{x_k}\rho^{(k)}(w)=\rho(w)+
[w,T \tilde \rho(w)],
\end{eqnarray}
where $w$ is the unknown $Q\times Q$ matrix function of the $N+1$
independent variables $(x_1,..,x_N,t)\in\RR^{N+1}$, $T$ is any
constant diagonal matrix, $[\cdot ,\cdot ]$ is the usual commutator
between matrices, and $\rho^{(k)},~\rho,~\tilde
\rho:~\RR\to\RR,~k=1,..,N$ are $N+2$ arbitrary scalar functions
representable by the positive power series:
 \begin{eqnarray}\label{ser_rho0}
 \rho^{(j)}(z)=\sum_{i\ge 0} \alpha^{(j;i)} z^i,\;\;\;
 \rho(z)=\sum_{i\ge 0} \alpha^{(i)} z^i,\;\;\;
 \tilde \rho(z)=\sum_{i\ge 0} \tilde \alpha^{(i)} z^i,
 \end{eqnarray}
 where $\alpha^{(j;i)}$, $\alpha^{(i)}$ and $\tilde \alpha^{(i)}$ are scalar constants, 
 so that the quantities
$\rho^{(k)}(w),~k=1,..,N$, $\rho(w)$ and $\tilde\rho(w)$ are
well-defined  functions of the matrix $w$. 

{\it Remark.} We  may always apply the following change of variable  $t$ and field $w$
\begin{eqnarray}
\partial_t + \sum_{j=1}^N \alpha^{(j;0)} \partial_{x_j} \to \partial_t,\;\;w\to e^{-\tilde \alpha^{(0)} T t } w e^{\tilde \alpha^{(0)} T t }
\end{eqnarray}
which is equivalent to putting 
\begin{eqnarray}\label{rho_10}
\alpha^{(j;0)} = 0,\;\;\;\; \tilde \alpha^{(0)} = 0 
\end{eqnarray}
in the eqs.(\ref{ser_rho0})
without loss of generality.

Solutions space to the matrix equation (\ref{equ-w}) with $\rho=0$ is implicitly described by the following algebraic  equation:
\begin{eqnarray}\label{w_C_G0}
&&w_{\alpha\beta}=\\\nonumber
&&
\sum_{\gamma=1}^Q \Big(e^{ -
 T_\alpha \tilde \rho(w) t
} 
 F_{\alpha\gamma}
\Big( x_1 I -  \rho^{(1)}(w)  t ,\dots,
x_N I- \rho^{(N)}(w)  t \Big) e^{ 
 T_\gamma \tilde \rho(w) t
} 
 \Big)_{\gamma\beta},
\\\nonumber
&&
\alpha=1,\dots,Q,\;\;\beta=1,\dots,  Q
,\;\;I\;\;{\mbox{is $Q\times Q$ identity matrix}}
.
\end{eqnarray}

If $\rho=0$, then eq.(\ref{equ-w}) has infinitely many commuting flows. In this case, it is convenient to replace eq.(\ref{equ-w}) by the next equation 
\begin{eqnarray}\label{equ-w_flows}\label{equ-w_com}
w_{t_m}+\sum\limits_{k=1}^N w_{x_k}\rho^{(mk)}(w)=
[w,T^m \tilde \rho^{(m)}(w)],\;\;m=1,2,\dots,
\end{eqnarray}
where index $m$ enumerates commuting flows, $\rho^{(mk)}$ are arbitrary scalar functions
representable by a positive power series
\begin{eqnarray}\label{ser_rho}
 \rho^{(mj)}(z)=\sum_{i\ge 0} \alpha^{(mj;i)} z^i,\;\;\;
 \tilde \rho^{(m)}(z)=\sum_{i\ge 0} \tilde \alpha^{(m;i)} z^i,
 \end{eqnarray}
$ \alpha^{(mj;i)}$, $\tilde  \alpha^{(m;i)}$ are scalar constants, $T^{(m)}$ are 
diagonal matrices.
Note, that the transposed  equation 
\begin{eqnarray}\label{equ-w_com_dual}
\tilde w_{t_m}+\sum\limits_{k=1}^N \rho^{(mk)}(\tilde w) 
\tilde w_{x_k}=
[\tilde \rho^{(m)}(\tilde w) T^m,\tilde w ],\;\;\tilde w=w^T,\;\;m=1,2,\dots
\end{eqnarray}
is also integrable.

Equation (\ref{equ-w}) as well as its generalizations whose solutions spaces may be completely described implicitly by the system of non-differential equations will be referred to as PDE0 in this paper for the sake of brevity.

Using either the algebraic manipulations (Sec.\ref{Algebraic}) or
the modified version of the dressing method (Sec.\ref{Dressing}),
we will show that
PDE0s (\ref{equ-w_com}) 
may be viewed as lower dimensional reductions of the classical Self-dual type
$S$-integrable PDEs. These PDEs are  referred to as 
PDE1s in this paper. 

\subsection{Family of PDE1s}
\label{Intr:Sint}
PDE1s may be viewed
as compatibility conditions of the following overdetermined  system of 
linear PDEs for some spectral 
function  $V(\lambda;x)$ (spectral system):
 \begin{eqnarray}\label{SP_s}
 &&
  V_{t_m}(\lambda;x) + \sum_{i=1}^N
 \lambda^{n_{mi}} V_{x_i}(\lambda;x) + 
 \sum_{i=1}^{\tilde N} \lambda^{\tilde n_{mi}} V(\lambda;x)T^{(i)} 
 =
 \sum_{i=0}^{N^{(m)}_{max}} \lambda^i V(\lambda;x) q^{(mi)}(x),
 \\\nonumber
 &&
 m=1,2.
\end{eqnarray}
Here  all functions are $Q\times Q$ matrix functions, $\lambda$ is a complex spectral parameter,
$n_{mi}$ and $\tilde n_{mi}$ are some integers, 
$N^{(m)}_{max}=\max(n_{mi_1},\tilde n_{mi_2},\;\;i_1=1,\dots,N,\;\;i_2=1
,\dots,\tilde N)$; $q^{(mi)}(x)$ are some matrix functions of 
$x=(x_1,x_2...,t_1,t_2...)$. Throughout this paper we use Greek letters in the lists of arguments for spectral parameters.
 The feature of this system is that its LHS is a first order differential  operator having coefficients which are  polynomial in $\lambda$ and independent  on the vector parameter $x$, i.e. $x$-dependent potentials appear only in the RHS, which is also polynomial in $\lambda$. 
Compatibility condition of any pair  of eqs.(\ref{SP_s}) results in  PDE1 for the functions $q^{(mi)}$. Note that eq.(\ref{SP_s}) uses only multiplication by the spectral parameter $\lambda$ and does not involve derivatives of the spectral function with respect to the spectral parameter. For this reason  such $S$-integrable PDEs as dispersion-less Kadomtsev-Petviashvili equation (dKP) and Heavenly equation may not be considered in the framework of the represented dressing algorithm. 

Remark, that the spectral system (\ref{SP_s}) may be replaced by the equivalent one
\begin{eqnarray}\label{fields_g}
&&
 \partial_{t_m} V(\lambda;x) + \sum_{i=1}^N A^{n_{mi}}(\lambda,\nu)*  \partial_{x_i}V(\nu;x) + \sum_{i=1}^{\tilde N}  A^{\tilde n_{mi}}(\lambda,\nu)* V(\nu;x)T^{(i)}=\\\nonumber
 &&
\sum_{i=0}^{N^{(m)}_{max}} A^i(\lambda,\nu)*V(\nu;x) q^{(mi)}(x),
\end{eqnarray}
where function $A(\lambda,\mu)$ replaces multiplication by  $\lambda$ and   we define operator $A^i$ as follows:
\begin{eqnarray}\label{AAA}
A^i=\underbrace{A*\cdots*A}_i.
\end{eqnarray}
Emphasise that $A$ is independent on $x$. The eq.(\ref{fields_g}) reduces to the eq.(\ref{SP_s})
 if 
 $A(\lambda,\nu)= \lambda\delta(\lambda-\mu)I$, where $I$ is $Q\times Q$ identity matrix.

\paragraph{Example 1: $N$-wave system.}

 Linear spectral system reads in the simplest case:
 \begin{eqnarray}\label{fields_g_s0}\label{lin_Nw}
 \partial_{t_m} V(\lambda;x)-A(\lambda,\nu) * V(\nu;x) T^{(m)} =V(\lambda;x) [T^{(m)}, w^{(0)}(x)],\;\;m=1,2.
\end{eqnarray}
Compatibility condition of this system yields the next PDE1:
\begin{eqnarray}\label{Hopf2}\label{nlin_Nw}
[T^{(1)}, w^{(0)}_{t_2}]-[T^{(2)} ,w^{(0)}_{t_1}] +
[[T^{(2)}, w^{(0)}],[T^{(1)} ,w^{(0)}]]=0.
\end{eqnarray}
 Most physical applications 
require reduction $w_{\alpha\beta}^{(0)}= \bar w_{\beta\alpha}^{(0)}$, where $\alpha\neq \beta$ and ''bar'' means complex conjugate.
This equation is well known  
as (1+1)-dimensional $N$-wave equation \cite{ZMNP}. We consider eq.(\ref{nlin_Nw}) without reductions. 

\paragraph{Example 2: Pohlmeyer equation.}
Another  example is assotiated with the following spectral system:
\begin{eqnarray}\label{fields_g_s}\label{lin_P}
 \partial_{t_m} V(\lambda;x) +  A(\lambda,\nu)*  \partial_{x_m}V(\nu;x) =V(\lambda;x) w^{(0)}_{x_m}(x),\;\;m=1,2.
\end{eqnarray}
Compatibility condition of this system yields the next PDE1:
\begin{eqnarray}\label{nlin_P}\label{P}
w^{(0)}_{x_1t_2}-w^{(0)}_{x_2t_1} =[w^{(0)}_{x_1},w^{(0)}_{x_2}],
\end{eqnarray}
which may be written in a different form:
\begin{eqnarray}\label{nlin_P_J}
&&
(J^{-1} J_{t_1})_{x_2}-(J^{-1} J_{t_2})_{x_1}=0,\\\nonumber \\\nonumber 
&&
w^{(0)}_{x_1}=J^{-1} J_{t_1},
\;\;\;\;
w^{(0)}_{x_2}=J^{-1} J_{t_2}.
\end{eqnarray}
This is Pohlmeyer equation \cite{P1,P2,P3}. Most applications  in Physics is assotiated with its reduction $J^+=\pm J$, $\det J=1$, which yields (Anty)Selfdual Yang-Mills equation (ASDYM). This equation has been studied by many authors:  \cite{BPST,BZ1,BZ2,AH,ADHM,DM,Ward0,Sanchez} (instanton solutions),
 \cite{Protogenov} (merons), \cite{Kr3,Korotkin} (finite gap solutions). (2+1)-dimensional version of ASDYM 
 assotiated with  the reduction
 \begin{eqnarray}\label{Jw_red}
\partial_{x_2}J=\partial_{t_1}J,\;\;\partial_{x_2}w^{(0)}=
\partial_{t_1}w^{(0)}
\end{eqnarray}
is also well known:
\begin{eqnarray}\label{P_red}
&&
w^{(0)}_{t_1 t_1}-w^{(0)}_{t_2 x_1} = [w^{(0)}_{t_1},w^{(0)}_{x_1}]\;\;\;\;\;\;{\mbox{or}}\\\nonumber
&&
(J^{-1} J_{t_1})_{t_1}-(J^{-1} J_{t_2})_{x_1}=0.
\end{eqnarray}
 It has been studied in \cite{ZM_sd,V} (Initial Value Problem), \cite{Ward1,Dimakis} (localized solutions). Here we will deal with eqs.(\ref{P}) and (\ref{P_red}a) without any reduction for the field $w^{(0)}$. 


\subsection{Basic novalties}
\label{Intr:nov}
There are  two manifolds of particular solutions to PDE1s. First manifold is assotiated with the uniquely solvable linear integral equation for some  spectral  function 
(Zakharov-Shabat method \cite{ZM0},  classical $\bar{\partial}$-problem \cite{Z0}). We concentrate on the  
second manifold of solutions  which  is described implicitly by a system of non-differential equations (see Sec.\ref{Algebraic}).
Simultaneously, this manifold is solution manifold to appropriate lower dimensional PDE0.
  Proper choice of initial data leads to wave profile breaking. Such solutions for Pohlmeyer equation will be discussed. 
  In turn, second manifold of solutions has a sub-manifold of explicit solutions.   Regarding (1+1)-dimensional $N$-wave equation, only such solutions from the second manifold are available.  It is likely that the same conclusion is valid for any other (1+1)-dimensional $S$-integrable model.

It is remarkable, that there is a version of the dressing method, which joins both manifolds of solutions to DPE1s, Sec.\ref{Dressing}. This fact is remarkable, because the second manifold is beyond the scope of the   classical dressing method. In addition, this fact demonstrates that  two significantly different classes of PDEs ($S$-integrable PDE1s and integrable by the method of characteristics PDE0s) are combined by the dressing method. 
There are examples of similar combinations.
For instance, dressing method  combining $S$- and $C$-integrable models has been suggested  in \cite{Z}; dressing method  combining $C$-integrable equations and equations integrable by the method of characteristics  has been proposed in \cite{ZS2}.   

All these examples tell us about flexibility of the dressing method which is important for development of the uniform method for integration of multidimensional nonlinear PDEs.
  
Usually,  we will use notations PDE1($t_{m_1},t_{m_2};w^{(0)}$),
 PDE0($t_m;n_0,w$) and PDE0($t_m;k_0,\tilde w$) with 
 arguments reflecting field as well as  derivative(s)  of field 
  with respect to  variable(s) $t_m$  appearing in  PDE. 
 Of course,  PDE0s and PDE1s involve also derivatives with respect to $x_i$. However, we do not represent them in the lists of arguments. 
Here $w(x)$ is  $n_0 Q\times n_0 Q$ matrix and $\tilde w(x)$ is $k_0 Q\times k_0 Q$ matrix block functions:
\begin{eqnarray}\label{wwww}\label{www}
w(x)&=& \left[\begin{array}{ccccc}
w^{(0)}(x) &w^{(1)}(x)&\cdots& w^{(n_0-2)}(x)&w^{(n_0-1)}(x)\cr
I &{\bf 0}&\cdots &{\bf 0}& r^{(1)} \cr
{\bf 0} &I&\cdots &{\bf 0}& r^{(2)} \cr
\vdots&\vdots&\vdots&\vdots&\vdots\cr
{\bf 0}&{\bf 0}&\cdots&I&r^{(n_0-1)}
\end{array}\right],\\\nonumber
\tilde w(x)&=& \left[\begin{array}{ccccc}
-\tilde w^{(0)}(x)&I&{\bf 0}&\cdots&{\bf 0} \cr
-\tilde w^{(1)}(x)&{\bf 0}&I&\cdots&{\bf 0}\cr 
\vdots&\vdots&\vdots&\vdots&\vdots\cr
-\tilde w^{(k_0-2)}(x)&{\bf 0}&{\bf 0}&\cdots&I\cr
-\tilde w^{(k_0-1)}(x)& \tilde r^{(1)} & \tilde r^{(2)}&\cdots &\tilde r^{(k_0-1)}
\end{array}\right],\;\;\tilde w^{(0)}=w^{(0)}, 
\end{eqnarray}
where $I$ and ${\bf 0}$ are $Q\times Q$ identity and zero matrices, $w^{(i)}$ and $\tilde w^{(i)}$ are $Q\times Q$ matrix functions, $n_0$ and $k_0$ are  arbitrary integers, $r^{(i)}$, $\tilde r^{(i)}$, $i=1,\dots,n_0-1$ are arbitrary constant  $Q\times Q$ matrices (if $T^{(i)} =0$ in eqs.(\ref{wdir},\ref{twdual})) or arbitrary  diagonal $Q\times Q$  matrices (if $T^{(i)} \neq 0$). 
Each PDE1($t_1,t_2; w^{(0)}$) written for the matrix field $w^{(0)}$ and involving derivatives with respect to $t_1$ and $t_2$  possesses a family of PDE0($t_m;n_0$,$w$) and PDE0($t_m;k_0$,$\tilde w$), $m=1,2$, $n_0,k_0=1,2,\dots$, as  lower dimensional reductions. Of course, PDE0($t_1;n_0$,$w$) is compatible with PDE0($t_2;n_0$,$w$), as well as PDE0($t_1;n_0$,$\tilde w$) is compatible with PDE0($t_2;n_0$,$\tilde w$).

For instance, in Sec.\ref{Algebraic} we will derive the following PDE0s assotiated with  $N$-wave equation (Sec.\ref{Sec:Nw_lin}): 
\begin{eqnarray}\label{INTR:w0n+1_320_w}
&&
w_{t_m} 
 -
  [w,T^{(m)}_{n_0}w] 
 =
 0,\\\nonumber
 &&
\tilde w_{t_m} 
 -
   [\tilde w \tilde T^{(m)}_{k_0},\tilde w]
 = 0,\;\;m=1,2,\\\nonumber
 &&\hspace{2cm}
 T^{(m)}_{n}={\mbox{diag}}(\underbrace{T^{(m)}\cdots T^{(m)}}_{n}),\;\;\;\tilde T^{(m)}_{n}=-T^{(m)}_{n}
 \end{eqnarray}
 and PDE0s assotiated with Pohlmeyer equation (Sec.\ref{Sec:P}): 
 \begin{eqnarray}
\label{INTR:w0n+1_320_0_w}
&&
w_{t_m} + w_{x_m}w
  =
 0,\\\nonumber
&&
\tilde w_{t_m} + \tilde w \tilde w_{x_m}
  =
 0,\;\;m=1,2.
  \end{eqnarray}
  If reduction (\ref{Jw_red}) is imposed, then   eq.(\ref{INTR:w0n+1_320_0_w}) reduces to
  \begin{eqnarray}
  \label{INTR:w0n+1_320_0_w_red}
&&
w_{t_1} + w_{x_1}w
  =
 0,\;\;\;
 w_{t_2} + w_{t_1}w
  =
 0,\\
 \nonumber
&&
\tilde w_{t_1} + \tilde w \tilde w_{x_1}
  =
 0,\;\;\;\tilde w_{t_2} + \tilde w \tilde w_{t_1}
  =
 0.
  \end{eqnarray} 
 General forms of PDE0($t_m; n_0,w$) and PDE0($t_m;k_0,\tilde w$), $m=1,2$, which are lower dimensional reductions of some PDE1($t_1,t_2;w^{(0)}$) are following:
 \begin{eqnarray}\label{wdir}
 &&
 w_{t_m} + \sum_{i=1}^N w_{x_i}\rho^{(mi)}(w) = \sum_{i=1}^{\tilde N} [w,T^{(i)}_{n_0}\tilde \rho^{(mi)}(w)],\;\;\;m=1,2,
 \\\label{twdual}
&&
\tilde w_{t_m} + \sum_{i=1}^N \rho^{(mi)}(\tilde w) \tilde w_{x_i} = \sum_{i=1}^{\tilde N} [\tilde \rho^{(mi)}(\tilde w) \tilde T^{(i)}_{k_0},\tilde w],\;\;\;m=1,2,
 \end{eqnarray}
 where functions $ \tilde \rho^{(mi)}$ are defined by the series similar to the eq.(\ref{ser_rho}):
 \begin{eqnarray}\label{ser_trho}
 \tilde \rho^{(mj)}(z)=\sum_{i\ge 0} \tilde \alpha^{(mj;i)} z^i,
 \end{eqnarray}
 and $\tilde\alpha^{(mj;i)}$ are scalar constants. Remarks on the origin of eqs.(\ref{wdir}) and (\ref{twdual}) will be  given in  Sec.\ref{Second}, see n.\ref{32},\ref{34} and n.\ref{322},\ref{342} respectively.
Eq.(\ref{twdual})
has the form of transposed equation (\ref{wdir}) with replacements
\begin{eqnarray}\label{repl}
 \tilde w= w^T,\;\;\;
 T^{(m)}_{n_0} \to \tilde T^{(m)}_{k_0}.
 \end{eqnarray}
 Because of this similarity, 
we will deal with PDE1($t_1,t_2;w^{(0)}$) and PDE0($t_m;n_0,w$), $m=1,2$. Regarding PDE0($t_m;k_0,\tilde w$), only some details 
will be given. 

 {\it Remark 1.} Although any PDE1 is compatibility condition of eqs.(\ref{fields_g}),  we are not able to represent  general form of PDE1s explicitly, unlike general form of PDE0, eqs.(\ref{wdir},\ref{twdual}).

{\it Remark 2.} The eq.(\ref{wdir}) defers from the
eq.(\ref{equ-w_com}) by the sum in the RHS. However, eq.(\ref{wdir}) may be treated  by the methods developed in \cite{SZ1,ZS2} for eq.(\ref{equ-w_com}).

{\it Remark 3.} The eqs.(\ref{INTR:w0n+1_320_w},\ref{INTR:w0n+1_320_0_w},\ref{wdir},\ref{twdual}) (where $m=1,2$) have infinitely many commuting flows corresponding to $m>2$ in these equations.

 \subsection{Brief contents} 
\label{Contents}
In the next section (Sec.\ref{Algebraic}) we  give algebraic 
description  of PDE1($t_1,t_2;w^{(0)}$), PDE0($t_m;n_0,w$) and PDE0($t_m;k_0,\tilde w$). Derivation of PDE1 (namely,  (1+1)-dimensional $N$-wave and (3+1)- and (2+1)- dimensional Pohlmeyer equations)  assotiated with 
appropriate PDE0($t_m;n_0,w$) and PDE0($t_m;k_0,\tilde w$), $m=1,2$ is given in Sec.\ref{PDE1_PDE0}. Assotiated linear overdetermined systems have been  described therein as well. 
 Using slightly modified results of \cite{SZ1,ZS2} we  represent the
non-differential matrix equations implicitly describing the solutions spaces  to the above PDE0s and assotiated solutions manifolds to the appropriate PDE1s, Sec.\ref{Solutions_alg}.  
We  derive some  manifolds of explicit solutions to $N$-wave and Pohlmeyer equations and describe solutions with  wave    
profile breaking for (2+1)- and (3+1)-dimensional Pohlmeyer equations.

A new version of the dressing method describing PDE0s and their relations with PDE1s is proposed in Sec.\ref{Dressing}. We  have derived examples of PDE0s and examples of PDE1s (following Sec.\ref{Algebraic},  $N$-wave  and Pohlmeyer equations are taken as examples of PDE1s) together with appropriate linear overdetermined systems (spectral systems), Sec.\ref{Section:M_1}.
Classical solutions manifolds to PDE1s (Sec.\ref{non-ev}) and solutions manifolds with  wave profile breaking to PDE1s and PDE0s (Sec.\ref{sol_PDE0}) have been  derived in Sec.\ref{Solutions}. 

Although this version of the dressing method is mostly straightforward in spirit of the derivation of PDE1s, it does not supply full solutions spaces to PDE0s, unlike the method of characteristics. 
The second version of the dressing algorithm exhibits fullness of the solutions spaces to PDE0s, Sec.\ref{Second}.

Conclusions are given in Sec.\ref{Conclusions}.


\section{Algebraic description of  PDE1s and PDE0s}
\label{Algebraic}
We consider three examples of 
PDE1($t_1,t_2;w^{(0)}$) (namely, eqs.(\ref{nlin_Nw}) and (\ref{nlin_P}) together with its (2+1)-dimensional reduction (\ref{P_red}a) ) and 
appropriate families of PDE0($t_m;n_0,w$) and PDE0($t_m;k_0,\tilde w$), $m=1,2$, in 
the framework of the algebraic approach, see 
Secs.\ref{Sec:Nw},\ref{Sec:P}. 
The solutions spaces to PDE1($t_1,t_2;w^{(0)}$)  and PDE0($t_m;n_0,w$), $m=1,2$,  will be 
 investigated in Sec.\ref{Solutions_alg}.

\subsection{Derivation of PDE1s and appropriate families of PDE0s}
\label{PDE1_PDE0}
\subsubsection{(1+1)-dimensional $N$-wave equation }
\label{Sec:Nw}
It is well known, that any $S$-integrable equation can be represented as  system of two commuting discrete chains. Mostly explicitly this representation  is given in Sato approach to  integrability, see for instance, \cite{Sato_KP}. However, any known dressing method gives rise to this representation. We  show, that  $N$-wave equation can be derived  from the 
 following pair of commuting discrete chains with two 
discrete parameters:
\begin{eqnarray}
\label{w0n+1}
&&w^{(kn)}_{t_m}  =w^{(k(n+1))} T^{(m)} 
-T^{(m)} w^{((k+1)n)}  +
  w^{(k0)} T^{(m)} w^{(0n)}  , \;\;m=1,2,
\end{eqnarray}
supplemented by the next non-differential relation among $w^{(ij)}$:
\begin{eqnarray}
\label{w0n+1_2_0}\label{w_discr}\label{w0n+1_2_alg}
&&
w^{((k+1)n)} =w^{(k(n+1))} + w^{(k0)} w^{(0n)}
\end{eqnarray}
(see also 
Sec.\ref{Char_Gen}, eqs.(\ref{w0n+1_2},\ref{w0n+12}) with 
$s^{(m)}=0$).
In fact, first of all remark, that eq.(\ref{w0n+1})  gives rise to  two alternative discrete chains with single discrete parameter in view of (\ref{w0n+1_2_0}). The first chain follows after  
putting $k=0$ and 
eliminating $w^{(1 n)}$ using eq.(\ref{w0n+1_2_0}):
\begin{eqnarray}\label{U_sp220_Nw}
&&
w^{(n)}_{t_m} -[ w^{(n+1)},T^{(m)}] + 
 [T^{(m)},w^{(0)}] w^{(n)}=
 0, \;\;m=1,2,
\end{eqnarray}
where 
\begin{eqnarray}\label{w_w}
w^{(n)}=w^{(0n)}, \;\;n=0,1,\dots.
\end{eqnarray}
The second discrete chain follows after putting $n=0$ and eliminating $w^{(k1)}$ from (\ref{w0n+1}) using (\ref{w0n+1_2_0}):
\begin{eqnarray}\label{U_sp220_Nw_tilde}
&&
\tilde w^{(k)}_{t_m} -[\tilde w^{(k+1)},T^{(m)}] + 
 \tilde w^{(k)} [\tilde w^{(0)},T^{(m)}]=
 0, \;\;m=1,2,
\end{eqnarray}
where 
\begin{eqnarray}\label{tilde_w_w}
\tilde w^{(k)}=w^{(k0)}, \;\;k=0,1,\dots,\;\;\tilde w^{(0)}=w^{(0)}.
\end{eqnarray}
Finally, in  order to write PDE1 for the function $w^{(0)}$,
we fix $n=0$ in (\ref{U_sp220_Nw}) and eliminate $w^{(1)}$, or
fix $k=0$ in (\ref{U_sp220_Nw_tilde}) and eliminate 
$\tilde w^{(1)}$.
In both cases we will end up with the same equation
(\ref{nlin_Nw}).

Along with PDE1($t_1,t_2;w^{(0)}$) (\ref{nlin_Nw}) 
we may derive  the family  of PDE0($t_m;n_0,w$), $m=1,2$, from the eqs.(\ref{U_sp220_Nw}) imposing the reduction
\begin{eqnarray}\label{red2}
w^{(n_0)}(x)=\sum_{i=0}^{n_0-1}w^{(i)}(x)r^{(i)},
\end{eqnarray}
where $n_0$ is an arbitrary integer and $r^{(i)}$ are arbitrary diagonal  constant matrices.
Similarly, we may derive PDE0($t_m;k_0,\tilde w$), $m=1,2$, from the eq.(\ref{U_sp220_Nw_tilde}) with the reduction
\begin{eqnarray}\label{red1}
\tilde w^{(k_0)}(x)=\sum_{i=0}^{k_0-1}\tilde r^{(i)}\tilde w^{(i)} (x),\;\;\tilde w^{(0)}= w^{(0)}, 
\end{eqnarray}
where  $k_0$ is arbitrary integer and $\tilde r^{(i)}$ are arbitrary diagonal  constant matrices. Both (\ref{red2}) and (\ref{red1})
are closures of the chains (\ref{U_sp220_Nw}) and (\ref{U_sp220_Nw_tilde}) respectively.
However, parameters $r^{(0)}$ and $\tilde r^{(0)}$ may be removed from the  PDE0s by the shifts of fields $w^{(n_0-1)}+r^{(0)} \to w^{(n_0-1)}$ and $-\tilde w^{(k_0-1)}+\tilde r^{(0)} \to -\tilde w^{(k_0-1)}$.
Thus hereafter we take 
\begin{eqnarray}\label{red2_r}
r^{(0)}=\tilde r^{(0)}=0
\end{eqnarray}
without loss of generality.
Parameters $n_0$, $k_0$, $r^{(i)}$ and $\tilde r^{(i)}$ appear in the definitions of $w$ and $\tilde w$, eqs.(\ref{wwww}).
The simplest explicit examples of PDE0($t_m;n_0,w$) and PDE0($t_m;k_0,\tilde w$) are following:  
\begin{eqnarray}\label{w0n+1_320}
 PDE0(t_m;1,w):&& 
w^{(0)}_{t_m} 
 + 
  [T^{(m)},w^{(0)}]w^{(0)} 
 =
 0,\;\;m=1,2,\\\label{w0n+1_32}
  PDE0(t_m;2,w),\;\;r^{(1)}=0:&&
w^{(0)}_{t_m} -[ w^{(1)},T^{(m)}]
  + 
  [T^{(m)},w^{(0)}]w^{(0)}=
 0,\\\nonumber
 && 
w^{(1)}_{t_m} 
  + 
  [T^{(m)},w^{(0)}]w^{(1)} =
 0, \;\;m=1,2,
\\\label{w0n+1_32_tilde}
PDE0(t_m;1,\tilde w):&&
\tilde w^{(0)}_{t_m}
  +
  \tilde w^{(0)}[\tilde w^{(0)},T^{(m)}]=
 0,\\\nonumber
PDE0(t_m;2,\tilde w) ,\;\;\tilde r^{(1)}=0:
&&
\tilde w^{(0)}_{t_m} -[ \tilde w^{(1)},T^{(m)}]
  +
 \tilde w^{(0)} [\tilde w^{(0)},T^{(m)}]=
 0,\\\nonumber
 && 
\tilde w^{(1)}_{t_m}
  +
 \tilde w^{(1)} [\tilde w^{(0)},T^{(m)}] =
 0,\;\; m=1,2.
\end{eqnarray}
It is not difficult  to observe that the above
  PDE0($t_m;n_0,w$)
and PDE0($t_m;k_0,\tilde w$)  with  arbitrary $n_0$ and $k_0$, may be written in the form 
(\ref{INTR:w0n+1_320_w}a) and (\ref{INTR:w0n+1_320_w}b) respectively,
 where $w$ and $\tilde w$ are defined by the   formulae (\ref{wwww}).


 Since the eq. (\ref{INTR:w0n+1_320_w}b) follows from the 
 eq.(\ref{INTR:w0n+1_320_w}a) after transposition with replacements (\ref{repl}),
  it is enough to study 
 one of them, say, eq.(\ref{INTR:w0n+1_320_w}a).
Slightly modifying result of  \cite{ZS2}, 
 we describe the solutions space to the eq.(\ref{INTR:w0n+1_320_w}a) with arbitrary 
 integer $n_0$ by the next implicit algebraic equation: 
\begin{eqnarray}\label{w_C}
&&
w_{\alpha\beta}= \sum\limits_{\gamma=1}^{n_0Q} \Big(  e^{-\sum\limits_{m=1}^2 
 (T^{(m)}_{n_0})_\alpha   w t_m}
 F_{\alpha\gamma}(w)\;\; e^{\sum\limits_{m=1}^2 
 ({T^{(m)}_{n_0}})_\gamma   w t_m}
 \Big)_{\gamma\beta},
\\\nonumber
&&
\alpha,\beta=1,\dots, n_0 Q,
\end{eqnarray}
where $F(z_0)$ is $n_0 Q\times n_0 Q$ matrix function of single argument.
Details of derivation of this formula by the dressing method are given in Secs.\ref{sol_PDE0}
and \ref{Second}.
However, the structure of $w$  tells us that 
 $F$ must have the following structure:
\begin{eqnarray}\label{F_0Nw}
 F_{\alpha\beta}(z_0)=\left\{\begin{array}{ll}
{\mbox{arbitrary scalar function of arguments}}, &\alpha\le Q\cr 
\delta_{\alpha\beta} z_0, &\alpha> Q
\end{array}\right.,
\end{eqnarray}
so that equation (\ref{w_C}) becomes an identity for $\alpha> Q$.
 Thus, the square matrix algebraic equation (\ref{w_C}) reduces to the rectangular matrix equation with $\alpha=1,\dots,Q$ and $\beta=1,\dots, n_0 Q$. 
  The scalar functions of single argument $F_{\alpha\beta}(z_0)$,  $\alpha =1,\dots,Q$, $\beta =1,\dots,n_0 Q$ are arbitrary.
The simplest case $n_0=1$ yields
$w\equiv w^{(0)}$, $T^{(m)}_1\equiv T^{(m)}$:
\begin{eqnarray}\label{w_C_0}
&&
w^{(0)}_{\alpha\beta}= \sum\limits_{\gamma=1}^Q \Big(  e^{-\sum\limits_{m=1}^2 
 T^{(m)}_\alpha   w^{(0)} t_m}
 F_{\alpha\gamma}(w^{(0)})\;\; e^{\sum\limits_{m=1}^2 
 T^{(m)}_\gamma   w^{(0)} t_m}
 \Big)_{\gamma\beta},\;\;\alpha,\beta=1,\dots,Q.
\end{eqnarray}

\paragraph{Assotiated linear overdetermined system of PDEs.}
\label{Sec:Nw_lin}
As it was mentioned above, the linear spectral problem for $N$-wave equation is given by  the eqs.(\ref{lin_Nw}), where $V(\lambda;x)$ is, generally speaking, a rectangular  $lQ\times Q$ matrix function, $l$ is some integer (to anticipate, $l$ is assotiated with the dimension of the kernel of the integral operator in eq.(\ref{u1}); $l=2$ in Secs.\ref{Section:M_1},\ref{Solutions}).
One can extend  this spectral system introducing discrete parameter $n$ by the following formulae:
\begin{eqnarray}\label{U_sp21_00_Nw} \label{V_nondif}\label{U_sp21_00} 
&&
A(\lambda,\nu)*V^{(n)}(\nu;x) = V^{(n+1)}(\lambda;x) +
V^{(0)}(\lambda;x)  w^{(n)}(x),
\\\label{PP1_Nw}
&&
V^{(n)}_{t_m}(\lambda;x) -
A(\lambda,\nu)*V^{(n)}(\nu;x)  T^{(m)}=
V^{(0)}(\lambda;x) [T^{(m)}, w^{(n)}(x)],\\\nonumber
&&
n=0,1,2,\dots, m=1,2.
\end{eqnarray}
which becomes eq.(\ref{lin_Nw}) if $n=0$ and $V\equiv V^{(0)}$. This extension can be derived formally, for instance, using 
a version of the dressing method developed in Sec.\ref{Dressing}, see Sec.\ref{Char_Gen} eq.(\ref{U_sp21}) with  $s^{(m)}=0$.

Let us  consider the reduction (\ref{red2},\ref{red2_r}) for the fields $w^{(n)}$, which  causes the appropriate reduction for the spectral functions $V^{(n)}$:
 \begin{eqnarray}
 \label{red2V}
 V^{(n_0)}=\sum_{i=1}^{n_0-1} V^{(i)} r^{(i)}.
 \end{eqnarray}
In view of this reduction, we may introduce the block-vector spectral function  
\begin{eqnarray}\label{V_block}
{\bf V}=[V^{(0)} \cdots V^{(n_0-1)}].
\end{eqnarray} 
Then eqs.(\ref{U_sp21_00_Nw},\ref{PP1_Nw}) can be written in a compact form:
 \begin{eqnarray}\label{U_sp21_0_Nw}  \label{AV}\label{U_sp21_0}
&&
A(\lambda,\nu)*{\bf V}(\nu;x) = {\bf V}(\lambda;x) w(x)
\\\label{U_sp212_Nw}
&&
{\bf V}_{t_m}(\lambda;x)  -A(\lambda,\nu)*{\bf V}(\nu;x) T^{(m)}_{n_0}=
 {\bf V}(\lambda;x) [T^{(m)}_{n_0}, w(x)]
,\;\;\;
m=1,2.
\end{eqnarray}
An important feature of this system is  the eq.(\ref{U_sp21_0_Nw}) which has no derivatives with respect to variables $t_i$ and $x_i$. Thus, $w(x)$ in the RHS of this equation may be treated as a matrix parameter.
For this reason, eq.(\ref{U_sp21_0_Nw}) may be explicitly solved, at least,  for diagonalizable matrix $w$ (more complex case, when $w$ is transformable to the Jordan form, is not considered here): 
\begin{eqnarray}\label{w_diag}\label{PEP}
&&w(x)=P(x) E(x) P^{-1}(x),\;\;\\\label{P_norm}
&&P_{\alpha\alpha}=1,
\end{eqnarray}
where $E$ is a diagonal  matrix of eigenvalues, $P$ is a matrix of eigenvectors normalized by the condition (\ref{P_norm}).

 Let $A$ be in the  following form: 
\begin{eqnarray}\label{A_delta}
&& \;\;\;A(\lambda,\nu)=\lambda\delta(\lambda-\nu)I_l,
\end{eqnarray}
where $I_l$ is $l Q\times lQ$ identity matrix.
 Then solution of  eq.(\ref{U_sp21_0_Nw}) reads:
\begin{eqnarray}\label{V_expl}
&&
{\bf V}(\lambda;x)P(x)=
\hat {\bf V}(x)\delta(\lambda I - E),
\end{eqnarray}
where 
\begin{eqnarray}\label{V_block2}
\hat {\bf V}=[\hat V^{(0)} \; \cdots \; \hat V^{(n_0-1)}]
\end{eqnarray}
Thus, ${\bf V}(\lambda;x) P(x)$ is delta-function of $\lambda$. This is an interesting fact and a  non-trivial reduction imposed on the spectral function ${\bf V}(\lambda;x)$. It is remarkable, that there is a dressing algorithm providing consistency of this reduction with eq.(\ref{U_sp212_Nw}), see Sec.\ref{Char_Gen}. 

There is an important remark.
{\it Remark:} The formula (\ref{V_expl}) means that
\begin{eqnarray}\label{V_inf}
{\bf V}(\lambda;x) \to 0 \;\;\;{\mbox{as}}\;\;\;\lambda \to \infty
\end{eqnarray}
for any  function $w(x)$. Such a behaviour of the spectral function is in
 contradiction with  the classical dressing, where the spectral function is not vanishing at infinity. For this reason the solutions described in this section are not observable in the framework of  ISTM.

\subsubsection{Pohlmeyer equation}
\label{Sec:P}
 One can  show that Pohlmeyer equation  may be derived  from the 
 following pair of   commuting discrete chains with two discrete parameters:
\begin{eqnarray}
\label{w0n+1_alg}
&&w^{(kn)}_{t_m} + w^{((k+1)n)}_{x_m}-w^{(k0)} w^{(0n)}_{x_m} =0,
\end{eqnarray}
supplemented by the eq.(\ref{w_discr})
(see also  Sec.\ref{Char_Gen}, eq.(\ref{w0n+1_2},\ref{w0n+12}) with $s^{(m)}=1$, $T^{(m)}=0$).
 First of all,  similar to the Sec.\ref{Sec:Nw}, eq.(\ref{w0n+1_alg})  gives rise to two alternative chains with single discreet parameter. The first chain appears after putting $k=0$ and  eliminating  $w^{(1n)}$ using eq.(\ref{w_discr}): 
\begin{eqnarray}\label{U_sp220_alg}
&&
w^{(n)}_{t_m} + w^{(n+1)}_{x_m}+w^{(0)}_{x_m}
 w^{(n)}=
 0,\;\;n=0,1,2,\dots,\;\;m=1,2,
 \end{eqnarray}
 where fields $w^{(n)}$ are given by the eq.(\ref{w_w}).
 The second discrete chain appears after putting $n=0$ and eliminating $w^{(k1)}$ using eq.(\ref{w_discr}):
 \begin{eqnarray}
 \label{U_sp220_alg_tilde}
&&
 \tilde w^{(k)}_{t_m} + \tilde w^{(k+1)}_{x_m}-\tilde w^{(k)}\tilde w^{(0)}_{x_m}
 =
 0,\;\;n=0,1,2,\dots,\;\;m=1,2,
\end{eqnarray}
 where fields $\tilde w^{(n)}$ are defined by the eq.(\ref{tilde_w_w}).
Finally, in order to write PDE1 for $w^{(0)}$, we put $n=0$ in eq.(\ref{U_sp220_alg}) and eliminate
$w^{(1)}$ or put $k=0$ in  eq.(\ref{U_sp220_alg_tilde})  and eliminate
$\tilde w^{(1)}$. In both cases we end up with  eq.(\ref{nlin_P}).
 
Along with  PDE1($t_1,t_2;w^{(0)}$) (\ref{nlin_P}) we may derive PDE0($t_m;n_0,w$), $m=1,2$, imposing the reduction (\ref{red2},\ref{red2_r}) to the eq.(\ref{U_sp220_alg}) or PDE0($t_m;k_0,\tilde w$), $m=1,2$,  imposing the
 reduction (\ref{red1},\ref{red2_r}) to the  eq.(\ref{U_sp220_alg_tilde}). 
The simplest examples are following:
\begin{eqnarray}\label{w0n+1_320_0}
 PDE0(t_m;1,w):&& 
w^{(0)}_{t_m} + w^{(0)}_{x_m}w^{(0)}
  =
 0,\;\;m=1,2,\\\nonumber
 PDE0(t_m;2,w),\;\; r^{(1)}=0:&& 
w^{(0)}_{t_m} + w^{(1)}_{x_m}+w^{(0)}_{x_m}w^{(0)}=
 0,\\\nonumber
 && 
w^{(1)}_{t_m} +w^{(0)}_{x_m}w^{(1)}=
 0,\;\;m=1,2,\\
\label{w0n+1_320_0_tilde}
 PDE0(t_m;1,\tilde w):&& 
\tilde w^{(0)}_{t_m} - \tilde w^{(0)}\tilde w^{(0)}_{x_m}
  =
 0,\;\;m=1,2,\\\nonumber
 PDE0(t_m;2,\tilde w),\;\; \tilde r^{(1)}=0:&& 
\tilde w^{(0)}_{t_m} + \tilde w^{(1)}_{x_m}-\tilde w^{(0)}\tilde w^{(0)}_{x_m}=
 0,\\\nonumber
 && 
\tilde w^{(1)}_{t_m} -\tilde w^{(1)}\tilde w^{(0)}_{x_m}=
 0,\;\;m=1,2.
\end{eqnarray}
It is not difficult  to observe that the above
 PDE0($t_m;n_0,w$)
and PDE0($t_m;k_0,\tilde w$) with arbitrary $n_0$ and $k_0$ may be written in the form 
(\ref{INTR:w0n+1_320_0_w}a) and (\ref{INTR:w0n+1_320_0_w}b) respectively,
 where $w$ and $\tilde w$ are defined by the   formulae (\ref{wwww}).

Similar to the  previous section, we represent the detailed consideration only to the
eq.(\ref{INTR:w0n+1_320_0_w}a). Its solutions space is described by the next non-differential equation
 \cite{SZ1}:
\begin{eqnarray}\label{w_C_G_P}
&&
w_{\alpha\beta}= \sum_{\gamma=1}^{n_0 Q} \Big( 
F_{\alpha\gamma}(w, x_1 I_{n_0}- w t_1 , x_2 I_{n_0}- w  t_2)
 \Big)_{\gamma\beta},
\\\nonumber
&&
\alpha,\beta=1,\dots, n_0 Q.
\end{eqnarray}
Details of derivation of this formula by the dressing method are given in Secs.\ref{sol_PDE0}
and \ref{Second}.
Here, similar to the $N$-wave equation,  $F$ must have the  structure predicted by the structure of $w$ given by the eq.(\ref{wwww}a): 
\begin{eqnarray}\label{F_0P}
 F_{\alpha\beta}(z_0, z_1,z_2)=\left\{\begin{array}{ll}
{\mbox{arbitrary scalar function of arguments}}, &\alpha\le Q\cr 
\delta_{\alpha\beta} z_0, &\alpha> Q
\end{array}\right.,
\end{eqnarray}
i.e. the square matrix equation (\ref{w_C_G_P}) reduces to the rectangular one with
$\alpha=1,\dots, Q$, $\beta=1,\dots, n_0 Q$  and  arbitrary scalar functions of three arguments $F_{\alpha\beta}(z_0,z_1,z_2)$.
The simplest case $n_0=1$ yields $w\equiv w^{(0)}$:
\begin{eqnarray}\label{w_C_G_P_0}
&&
w^{(0)}_{\alpha\beta}= \sum_{\gamma=1}^Q \Big( 
F_{\alpha\gamma}(w^{(0)}, x_1 I- w^{(0)} t_1 , x_2 I- w^{(0)} t_2)
 \Big)_{\gamma\beta},
\alpha,\beta=1,\dots,Q.
\end{eqnarray}
Let  reduction (\ref{Jw_red}) be used. Consider PDE0($t_m;n_0,w$) (\ref{INTR:w0n+1_320_0_w_red}), $m=1,2$,  corresponding to PDE1($t_1,t_2;w^{(0)}$) (\ref{P_red}). 
Its solutions space is implicitly described by the next algebraic equation
\begin{eqnarray}\label{w_C_G_Pred}
&&
w_{\alpha\beta}= \sum_{\gamma=1}^{n_0 Q} \Big( 
F_{\alpha\gamma}(w,x_1 I_{n_0} - w  t_1 + w^2  t_2)
 \Big)_{\gamma\beta},
\\\nonumber
&&
\alpha=1,\dots,Q,\;\;\beta=1,\dots, n_0 Q
\end{eqnarray}
 with arbitrary scalar functions of two arguments $F_{\alpha\beta}(z_0,z_1)$.
The simplest case $n_0=1$ yields ($w\equiv w^{(0)}$):
\begin{eqnarray}\label{w_C_G_Pred_0}
&&
w^{(0)}_{\alpha\beta}= \sum_{\gamma=1}^Q \Big( 
F_{\alpha\gamma}( w^{(0)},x_1 I - w^{(0)} t_1 + (w^{(0)})^2  t_2)
 \Big)_{\gamma\beta},
\;\;\alpha,\beta=1,\dots,Q.
\end{eqnarray}

\paragraph{Assotiated linear overdetermined system.}
\label{Sec:P_lin}
The classical overdetermined system for Pohlmeyer equation 
is given by the eq.(\ref{lin_P}), 
where $V$ is $l Q\times Q$ matrix function, $l$ is some integer.
System (\ref{lin_P}) may be extended as follows (such extension may be derived, for instance, using a version of the dressing method developed in Sec.\ref{Dressing}, see Sec.\ref{Char_Gen} eq.(\ref{U_sp21}) with  $s^{(m)}=1$, $T^{(m)}=0$):
\begin{eqnarray}
\label{PP1}
&&\hspace{-3cm}
V^{(n)}_{t_m}(\lambda;x) +  A(\lambda,\nu) *V^{(n)}_{x_m}(\nu;x) -
 V^{(0)}(\lambda;x)  w^{(n)}_{x_m}(x) =
0,\\\nonumber
&&
m=1,2,
\end{eqnarray}
where $V^{(n+1)}$ is related with $V^{(n)}$ by the eq.(\ref{U_sp21_00}). 
Eq.(\ref{PP1}) becomes  eq.(\ref{lin_P}) if $n=0$ and $V\equiv V^{(0)}$. 

We consider the reduction (\ref{red2},\ref{red2_r},\ref{red2V})  and
 introduce the  block-vector spectral function (\ref{V_block}), which allows us to 
  write the eq.(\ref{PP1}) in a compact form:
 \begin{eqnarray}
&&
{\bf V}_{t_m}(\lambda;x) +  A(\lambda,\nu) *{\bf V}_{x_m}(\nu;x) 
-
 {\bf V}(\lambda;x)  w_{x_m}(x) =
0,\;\;\;
m=1,2,
\end{eqnarray}
while eq.(\ref{U_sp21_00}) becomes  
eq.(\ref{AV}). Thus, the spectral equation (\ref{AV}) appears in both spectral systems corresponding  to  $N$-wave  and Pohlmeyer equations. This spectral equation is universal equation  assotiated with reduction (\ref{red2V}).
Suppose that the matrix $w$ is diagonalizable and  representable in the form  (\ref{w_diag}). 
 Let $A(\lambda,\nu)$ be in the form (\ref{A_delta}).
 Then ${\bf V}(\lambda;x)$ is given by the eq.(\ref{V_expl}).
 The dressing algorithm  consistent with  reductions (\ref{red2},\ref{red2_r},\ref{red2V}) will be represented in Secs.\ref{Char_Gen} and \ref{Second}.

\subsection{Solutions space}
\label{Solutions_alg}
As before, we assume diagonalizability of $w$, i.e. $w$ may be represented in the form (\ref{w_diag}),
and describe two different types of solutions to PDE0($t_m;n_0,w$), $m=1,2$, and assotiated solutions to PDE1($t_1,t_2;w^{(0)}$):
\begin{enumerate}
\item
All eigenvalues of $w$  are independent on $x$. This leads to the explicit  solutions $w$ without the wave profile breaking. In particular, the solutions in the form of rational function of exponential functions with linear in $x$ arguments are   available.  Solitary waves  belong to this type.
\item
Some or all eigenvalues  of $w$  depend on $x$. This leads to the solutions $w$ with   wave profile breaking.  Such specific behaviour is  typical for the  systems of hydrodynamic type.
\end{enumerate}
In this section  we will use the description of the solutions space 
to PDE0s given in \cite{SZ1}, i.e., instead of the equations 
(\ref{w_C},\ref{w_C_G_P},\ref{w_C_G_Pred}) we will consider the appropriate
non-differential equations describing the solutions spaces to the first 
order quasilinear  PDEs for $E$ and $P$. For  PDE0($t_m;n_0,w$) written in general form 
 (\ref{wdir}) these quasilinear PDEs read:
\begin{eqnarray}\label{w_eigE}
 &&
 E_{t_m} + \sum_{i=1}^N E_{x_i}\rho^{(mi)}(E) = 0,\;\;\;m=1,2,
\\\label{w_eigP}
 &&
 P_{t_m} + \sum_{i=1}^N P_{x_i}\rho^{(mi)}(E) = 
 \sum_{i=1}^{\tilde N} [P,T^{(i)}_{n_0}]\;\tilde \rho^{(mi)}(E),\;\;\;m=1,2.
\end{eqnarray}
This system is solvable by the method of characteristics \cite{SZ1} giving the following non-differential equations for $E_\beta$ and $P_{\alpha\beta}$:
\begin{eqnarray}\label{w_eigE_expl}
E_\beta &=&F^{(E)}_\beta\left(E_\beta,x_1-\sum_{m=1}^2 \rho^{(m1)}(E_\beta)t_m,\dots, x_N-\sum_{m=1}^2 \rho^{(mN)}(E_\beta)t_m\right),\;\;\\\label{w_eigP_expl}
P_{\alpha\beta}&=& e^{\displaystyle -\sum_{m=1}^2\sum_{j=1}^{\tilde N} (T^{(j)}_{n_0})_\alpha \tilde \rho^{(mj)}(E_\beta) t_m}
 F^{(P)}_{\alpha\beta} 
 \left(E_\beta,x_1-\sum_{m=1}^2 \rho^{(m1)}(E_\beta)t_m,\dots,\right. \\\nonumber
 &&
\left. x_N-\sum_{m=1}^2 \rho^{(mN)}(E_\beta)t_m\right)\;\
e^{\displaystyle \sum_{m=1}^2\sum_{j=1}^{\tilde N} (T^{(j)}_{n_0})_\beta \tilde \rho^{(mj)}(E_\beta) t_m},\;\;\alpha,\beta=1,\dots,n_0 Q,
\end{eqnarray}
where $F^{(E)}_\beta$ and $F^{(P)}_{\alpha\beta}$  are arbitrary functions of arguments if only there is no restriction on the structure of $w$.
 However, unlike \cite{SZ1,ZS2}, we
have the particular (rather then general) matrix structure of $w$, 
which requires the appropriate particular matrix structure of $P$. 
In fact,  eq.(\ref{PEP}) may be written in the form
\begin{eqnarray}\label{wPEP}
\sum_{\gamma=1}^{n_0Q} w_{\alpha\gamma} P_{\gamma_\beta}=P_{\alpha\beta}E_\beta,\;\;\alpha,\beta=1,\dots,n_0Q.
\end{eqnarray}
Eqs.(\ref{wPEP}) with $\alpha=1,\dots,Q$ and $\beta=1,\dots,n_0 Q$ should be taken as definitions of $w^{(j)}$, $j=0,\dots,n_0-1$ in terms of $E$ and $P$.
Other equations of the system (\ref{wPEP}) yield the next linear relations among the elements of $P$:
\begin{eqnarray}\label{lin_PP}
&&
\sum_{\gamma=1}^{n_0 Q} w_{\alpha\gamma} P_{\gamma\beta} = P_{\alpha\beta} E_\beta\;\;\;\Rightarrow
\;\;\;P_{(\alpha-Q)\beta}+\sum_{\gamma=(n_0-1)Q}^{n_0 Q}r^{(s+1)}_{\tilde \alpha \gamma} P_{\gamma\beta}=P_{\alpha\beta}E_\beta,\\\nonumber
&&
s=\left[\frac{\alpha}{Q}\right],\;\;\;\tilde \alpha=\alpha-s Q,
\;\;\;\alpha=Q+1,\dots,n_0Q,\;\;\beta=1,\dots,n_0Q,
\end{eqnarray}
where $[\cdot]$ means an integer part of number.
Eq.(\ref{lin_PP}) 
with $E_\beta\neq 0$ and arbitrary $r^{(i)}$ 
may be always solved for the following set of elements $P_{\alpha\beta}$:
\begin{eqnarray}\label{lin_PP_var}\label{PPP}
{\cal{P}}=\{P_{(\alpha-Q) \beta},P_{\alpha(\alpha-Q)};\;\alpha=Q+1,\dots,n_0 Q,\beta=1,\dots,n_0Q, \;\beta\neq \alpha-Q \}.
\end{eqnarray}
Other $n_0^2Q^2-(n_0-1)n_0Q^2 -n_0Q=n_0Q(Q-1)$  elements of $P$  are defined by the eq.(\ref{w_eigP_expl}) with arbitrary scalar functions $F^{(P)}_{\alpha\beta}$.
The case $n_0>1$, $E_{\beta_0}=0$ for some $\beta_0$ results in  some restrictions for $r^{(i)}$ so that  solution to the eqs.(\ref{lin_PP}) will differ from ${\cal{P}}$.
This case will not be considered.

All in all, the following algorithm gives either the explicit 
solutions to PDE0($t_m;n_0,w$) and assotiated PDE1($t_1,t_2;w^{(0)}$) ($E(x)=const$)  or the non-differential equations ($E(x)\neq const$) describing 
the solutions spaces  to PDE0($t_m;n_0,w$), $m=1,2$, and 
assotiated   PDE1($t_1,t_2;w^{(0)}$):
\begin{enumerate}
\item
Fix the arbitrary eigenvalues $E_\beta\neq 0$, $\alpha=1,\dots,n_0Q$, in the form (\ref{w_eigE_expl}).
\item
Fix $n_0 Q(Q-1)$  elements of $P_{\alpha\beta}\notin {\cal{P}}$ in the form (\ref{w_eigP_expl}).
\item
Find other elements of $P\in {\cal{P}}$ solving the linear system (\ref{lin_PP}) with normalization (\ref{P_norm}).
\item
Find $w$ using formula (\ref{PEP}).
\end{enumerate}
We will describe some solutions to $N$-wave and Pohlmeyer equations in
 Secs.(\ref{Solution_Nwave}) and (\ref{Solutions_P}) 
 respectively.
 
 Implicit representation of the solutions space by
  eqs.(\ref{w_eigE_expl},\ref{w_eigP_expl}) reveals solutions with wave profile breaking. 
  This phenomenon is assotiated with $E_\beta(x)\neq const$ (Pohlmeyer equation).
   We refer to the  break points manifold related with $E_\beta$ as ${\cal{E}}_\beta$. 
   We will show that break points manifold for $P_{\alpha\beta}$ is also ${\cal{E}}_\beta$.
   Then the break points manifold corresponding to $w$ will be $\displaystyle{\cal{E}}=\cup_{\beta=1}^{n_0 Q} {\cal{E}}_\beta$.
   
\subsubsection{$N$-wave equation} 
\label{Solution_Nwave}
The  equations (\ref{w_eigE},\ref{w_eigP}) corresponding to the  eq.(\ref{INTR:w0n+1_320_w}a)
 read:
\begin{eqnarray}
&&
E_{t_m}=0,\\\nonumber
&&
P_{t_m}=[P,T^{(m)}]E,\;\;\;m=1,2.
\end{eqnarray}
Then
eq.(\ref{w_C}) describing the solutions space to $N$-wave equation
will be replaced by the next pair of equations (the appropriate variant of the eqs.
(\ref{w_eigE_expl},\ref{w_eigP_expl})):
\begin{eqnarray}\label{Pe}
&&
E_\alpha=\epsilon_\alpha = const,\;\;\;\alpha=1,\dots,n_0Q,\\\nonumber
&&
P_{\alpha\beta}=e^{-\sum\limits_{m=1}^2 \big(T^{(m)}_{n_0}\big)_\alpha \epsilon_\beta t_m }(P_0)_{\alpha\beta} e^{\sum\limits_{m=1}^2 \big(T^{(m)}_{n_0}\big)_\beta \epsilon_\beta t_m },\\\nonumber
&&
\alpha=1,\dots,Q,\;\;\;\beta=1,\dots,n_0 Q
\end{eqnarray}
 where ${P_0}_{\alpha\beta}$ are  constants, related by the eq.(\ref{lin_PP}). Only $n_0Q(Q-1)$ of them may be arbitrary.
    Appropriate
 function $w^{(0)}$ will be rational function of exponents. 
 
 An important problem  is to exhibit such solutions which have no singularities in  domains of all independent variables. Two examples are given below:
 \begin{enumerate}
 \item Let $\epsilon_\alpha=i e_\alpha$, $e_\alpha$ be real parameters, and require
   $\det P \neq 0$, $t_1\in \RR$, $t_2 \in \RR$. 
 For instance let $Q=3$, $n_0=1$, ${P_0}_{23}={P_0}_{32}=0$, other  elements of $P_0$ equal $i$, 
$i^2=-1$ (${P_0}_{\alpha\beta}= i$),
 $T^{(n)}_\alpha=\alpha^n$, $\epsilon_\alpha=i\; (\alpha-1)$. 
 Corresponding $P$ reads:
 \begin{eqnarray}
 P=\left[\begin{array}{ccc}
1&i e^{i(t_1+3 t_2)}& i e^{4 i (t_1+4 t_2)}\cr
i &1&0\cr
i&0&1
 \end{array}\right],
 \end{eqnarray}
 which produces periodic $w^{(0)}$.
 \item Let $\epsilon_\alpha$ be real  parameters, $\det P \neq 0$, $t_1\in \RR$, $t_2 \in \RR$. However, unlike the previous example,  this is not enough to obtain the bounded  solutions, because one has to eliminate the exponential growth of $w$. Let  $Q=3$, 
 $n_0=1$. Exponential growth of $w^{(0)}$ disappears  
if one of the eigenvalues is zero and two others have opposite signs. 
 For instance, let $\epsilon_2=0$, $\epsilon_1=-\epsilon_3=1$,
 $T^{(n)}_\alpha=\alpha^n$, ${P_0}_{23}={P_0}_{32}=0$, other ${P_0}_{\alpha\beta}$ equal $i$. 
 Then $P$ reads:
 \begin{eqnarray}
 P=\left[\begin{array}{ccc}
1&i&i e^{-2(t_1+4 t_2)}\cr
 i e^{-  (t_1+3 t_2)}
&1&i e^{-(t_1+5 t_2)}\cr
i e^{-2(t_1+4 t_2)}&i&1
 \end{array}\right],
 \end{eqnarray}
 which produces the solitary wave  solution $w^{(0)}$.
 \end{enumerate}
An open problem remains how to provide  physically important 
  reduction $w_{\alpha\beta}=\bar w_{\beta\alpha}$.

 \subsubsection{Pohlmeyer equation}
 \label{Solutions_P}
 Eqs.(\ref{w_eigE},\ref{w_eigP}) corresponding to the eq.(\ref{INTR:w0n+1_320_0_w}a) read: 
 \begin{eqnarray}\label{E_P}
&&
E_{t_n}+E_{x_n} E=0,\\\nonumber
&&
P_{t_n}+P_{x_n} E =0,\;\;n=1,2.
\end{eqnarray}
The  general solution of the eqs.(\ref{E_P}) is described by the next pair of non-differential equations (the appropriate variant of the formulae (\ref{w_eigE_expl},\ref{w_eigP_expl})):
  \begin{eqnarray}\label{sol_P}
E_\beta &=& F^{(E)}_\beta(E_\beta,x_1-t_1 E_\beta, x_2-t_2 E_\beta),\\\nonumber
P_{\alpha\beta} &=& F^{(P)}_{\alpha\beta}(E_\beta, x_1-t_1 E_\beta, x_2-t_2 E_\beta),
\\\nonumber
&&
\alpha=1,\dots,Q,\;\;\beta=1,\dots,n_0 Q,
  \end{eqnarray}
  where $F^{(P)}_{\alpha\beta}(z_0,z_1,z_2)$ are related by the system (\ref{lin_PP}), so that only $n_0 Q(Q-1)$ of them remain arbitrary functions of three arguments.
 Similarly, eqs.(\ref{w_eigE},\ref{w_eigP}) corresponding to the 
 eq.(\ref{INTR:w0n+1_320_0_w_red}a) read: 
 \begin{eqnarray}
&&
E_{t_1}+E_{x_1} E=0,\;\;\;E_{t_2}+E_{t_1} E=0,\\\nonumber
&&
P_{t_1}+P_{x_1} E =0,\;\;\;P_{t_2}+P_{t_1} E =0.
\end{eqnarray}
 Solution to this system is implicitly described by the next pair of non-differential equations: 
  \begin{eqnarray}\label{sol_P_red}
E_\beta &=& F^{(E)}_\beta(E_\beta, x_1-t_1 E_\beta+t_2 E_\beta^2),\\\nonumber
P_{\alpha\beta} &=& F^{(P)}_{\alpha\beta}( E_\beta,x_1-t_1 E_\beta+t_2 E_\beta^2),
\\\nonumber
&&
\alpha=1,\dots,Q,\;\;\beta=1,\dots,n_0 Q,
  \end{eqnarray} 
  where $F^{(P)}_{\alpha\beta}(z_0,z_1)$ are related by the system (\ref{lin_PP}), so that only $n_0 Q(Q-1)$ of them remain arbitrary functions of two arguments.
  System (\ref{sol_P_red}) corresponds to the particular choice of $F^{(E)}$ and $F^{(P)}$ in (\ref{sol_P}). 
  Namely,  
  \begin{eqnarray}\label{redred}
  F^{(E)}(z_0,z_1,z_2)= F^{(E)}(z_0,z_1-z_0z_2),\;\;
  F^{(P)}(z_0,z_1,z_2)= F^{(P)}(z_0,z_1-z_0z_2),\;\;x_2=0.
  \end{eqnarray}
  
 Hereafter  we consider the "reduced" formulae (\ref{sol_P},\ref{sol_P_red}) with 
 \begin{eqnarray}\label{less_arg}
&&
 F^{(E)}(z_0,z_1,z_2)=F^{(E)}(z_1,z_2),\;\;
 F^{(P)}(z_0,z_1,z_2)=F^{(P)}(z_1,z_2),\\\nonumber
 &&
 F^{(E)}(z_0,z_1)=F^{(E)}(z_1),\;\;
 F^{(P)}(z_0,z_1)=F^{(P)}(z_1).
 \end{eqnarray}
 
 \paragraph{On explicit solutions.}
 As it was mentioned above, the explicit solutions correspond to $E_\beta =const$. 
 Then the matrix $P$ is the only object introducing dependence on $x$ in the field $w$. Let $F^{(P)}_{\alpha\beta}(z)$ be some smooth function vanishing as $|z|\to \infty$  with single maximum at $z=z^{(max)}$. The algebraic equations (\ref{sol_P_red})   describe two-dimensional surfaces in 3-dimensional space: 
 \begin{eqnarray}
 x_1-t_1 E_\beta + t_2 E_\beta^2=z^{(max)},\;\;\;\beta=1,\dots,n_0 Q.
 \end{eqnarray}
 Let $F^{(P)}_{\alpha\beta}(z_1,z_2)$ be smooth functions vanishing as $|z_1|\to \infty$ and $|z_2|\to\infty$ with single maximum at $z_i=z_i^{(max)}$, $i=1,2$. Algebraic equations (\ref{sol_P})   describe two-dimensional surfaces 
  in  4-dimensional space:
 \begin{eqnarray}
 x_1-t_1 E_\beta=z_1^{(max)},\;\;\;x_2 - t_2 E_\beta=z_2^{(max)},\;\;\;\beta=1,\dots,n_0 Q.
 \end{eqnarray}
 
 \paragraph{Break  points manifolds for wave (\ref{sol_P_red}).}
 Consider the  solutions  (\ref{sol_P_red}a) with  wave profile breaking assuming that $t_1$ is a physical  time and the space variables are $(x_1,t_2)$. The break points manifold ${\cal{E}}_\beta$ is defined by the two requirements. 
 
 The first requirement is that the derivative of $E_\beta$ in some direction(s) in space $(x_1,t_2)$ tends to infinity. 
 Calculating the partial derivatives
 \begin{eqnarray}\label{part_der}
\partial_{x_1} E_\beta &=&
\frac{\partial_z F^{(E)}_\beta(z)|_{z=x_1-t_1 E_\beta +t_2 E_\beta^2}}{d},\;\;\partial_{t_2} E_\beta =
\frac{E_\beta^2 \partial_z F^{(E)}_\beta(z)|_{z=x_1-t_1 E_\beta +t_2 E_\beta^2}}{d},\\\nonumber
&&
d=\left.\Big(1+(t_1-2 E_\beta t_2) \partial_{z}F^{(E)}_\beta(z) \Big)\right|_{z=x_1-t_1 E_\beta +t_2 E_\beta^2}
\end{eqnarray}
we see that all of them tend to infinity when the denominator  of the above expressions is zero:
 \begin{eqnarray}\label{first_req}
 \left.\Big(1+(t_1-2 E_\beta t_2) \partial_{z}F^{(E)}_\beta(z) \Big)\right|_{z=x_1-t_1 E_\beta +t_2 E_\beta^2}=0.
 \end{eqnarray}
 This condition tells us that $d=1\neq 0$ on the surface $t_1-2 E_\beta t_2=0$, so that the only 
way to have the infinite derivative of $E_\beta$ on this surface  is through the infinite derivative of the given function $F^{(E)}(z)$ in accordance with (\ref{part_der}). To eliminate this case, hereafter we consider only those functions $F^{(E)}(z)$ which have smooth behaviour and finite derivatives for $z\in\RR$. Thus, the necessary condition for the  break points manifold is 
\begin{eqnarray}\label{necessary}
t_1-2 E_\beta t_2 \neq 0.
\end{eqnarray} 
Note that the derivative of $E_\beta$ is zero in the next direction:
 \begin{eqnarray}
 \vec t \cdot \nabla E_\beta =0,\;\;\vec t= \frac{1}{\sqrt{1+E_\beta^4}}(E_\beta^2,-1),\;\; \nabla=(\partial_{x_1},\partial_{t_2}).
 \end{eqnarray}
The second requirement to the  break points manifold is that  function $E_\beta(x_1,t_1,t_2)$ must change its concavity on this manifold. This requirement is evident in (1+1)-dimensional  case  and is well described in  textbooks \cite{Whitham}. 
It is remarkable, that our (2+1)-dimensional  case is very similar.
To show this, we follow the classical strategy and take $E_\beta$ as independent variable in  the neighborhood of the breaking point. There are 
 two possible variants:
\begin{eqnarray}\label{Fz_x1}
1.&&{\mbox{$x_1$ is a function of $E_\beta$, $t_2$ and $t_1$}},
\\\label{Fz_t1}
2.&&{\mbox{$t_2$ is a function of $E_\beta$, $x_1$ and $t_1$}}.
\end{eqnarray}
Consider the  case (\ref{Fz_x1}).
All partial derivatives of $x_1$  are finite in the  break points manifold
(we do not need time-derivative for this analysis):
 \begin{eqnarray}
 &&
 \partial_{E_\beta} x_1 =\left. \frac{1+(t_1-2  t_2E_\beta) \partial_{z}F^{(E)}_\beta(z)}{\partial_{z}F^{(E)}_\beta(z)} \right|_{z=x_1-t_1 E_\beta +t_2 E_\beta^2},\\\nonumber
 &&
 \partial_{t_2} x_1 =-E_{\alpha}^2.
 \end{eqnarray}
The requirement (\ref{first_req}) is equivalent to $\partial_{E_\beta} x_1=0$. 
 Similarly, the second derivatives of $x_1$ are following:
 \begin{eqnarray}
 &&
 \partial_{E_\beta}^2 x_1 =-\left. \frac{2  t_2 (\partial_{z}F^{(E)}_\beta(z))^3+\partial_{z}^2F^{(E)}_\beta(z)
 }{(\partial_{z}F^{(E)}_\beta(z))^3} \right|_{z=x_1-t_1 E_\beta +t_2 E_\beta^2},\\\nonumber
 &&
 \partial_{t_2}\partial_{E_\beta} x_1 =-2E_\beta
 ,\\\nonumber
 &&
 \partial_{t_2}^2 x_1 =0.
 \end{eqnarray}
 We take equation
 $\partial_{E_\beta}^2 x_1=0$ as the
  second requirement to the   break points manifold, which, in view of (\ref{first_req}), reads:
 \begin{eqnarray}\label{two_br_cond}\label{br}
\left.\Big( 2  t_2 -(t_1-2 t_2 E_\beta)^3\partial_{z}^2F^{(E)}_\beta(z)\Big)\right|_{z=x_1-t_1 E_\beta +t_2 E_\beta^2}=0.
 \end{eqnarray}
 Eqs.(\ref{first_req},\ref{br}) are supplemented by the 
  eq.(\ref{sol_P_red}a).
  
  It is remarkable, that, starting with eq.(\ref{Fz_t1}), we will end up with the same equations (\ref{first_req},\ref{br}). We do not represent details of appropriate  calculations.
  
 The system (\ref{sol_P_red}a,\ref{first_req},\ref{two_br_cond}) can be solved explicitly:
 \begin{eqnarray}\label{br_expl}
{\cal{E}}_\beta: && t_2=-\left.
 \frac{\partial_z^2 F^{(E)}_\beta(z)}{2\Big( \partial_z F^{(E)}_\beta(z)\Big)^3}
 \right|_{z=\tilde F^{(E)}_\beta(E_\beta)},\\\nonumber
 &&
 t_1=- \left.\frac{\Big(\partial_zF^{(E)}_\beta(z)\Big)^2 + 
 E_\beta \partial_z^2 F^{(E)}_\beta(z)  }{\Big(\partial_z F^{(E)}_\beta(z)\Big)^3}
 \right|_{z=\tilde F^{(E)}_\beta(E_\beta)},\\\nonumber
 &&
 x_1=t_1 E_\beta - t_2 E_\beta^2 + \tilde F^{(E)}_\beta(E_\beta),
 \end{eqnarray}
 where $\tilde F^{(E)}_\beta(z)$ is inverse of $F^{(E)}_\beta(z)$: $F^{(E)}_\beta\Big( 
 \tilde F^{(E)}_\beta(z)\Big)=z$. This system of three equations describes 
the  break points manifold ${\cal{E}}_\beta$.
  We can calculate two-component velocity $(\partial_{t_1}x_1,\partial_{t_1} t_2)$ of the  break points manifold differentiating the eqs.(\ref{sol_P_red}a,\ref{first_req},\ref{two_br_cond}) 
 with respect to $t_1$ and assuming that variables $E_\beta$, $x_1$ and $t_2$ are functions of $t_1$:
 \begin{eqnarray}\label{vel}
 &&
 \partial_{t_1} E_\beta=\frac{2E_\beta t_2 -t_1}{E_\beta\Big(
 12 t_2^2 + (t_1-2E_\beta t_2 )^5 \partial_z^3 F(z)|_{z=x_1-E_\beta t_1 + E_\beta^2 t_2}\Big)},\\\nonumber
 &&
 \partial_{t_1} x_1 =\frac{E_\beta}{2} ,\;\;\;
 \partial_{t_1} t_2 = \frac{1}{2 E_\beta}
 \end{eqnarray}
 As a simple example, let 
 \begin{eqnarray}
 F^{(E)}_\beta(z)=-\tanh(z).
 \end{eqnarray}
 Then the systems (\ref{br_expl},\ref{vel}) yield:
 \begin{eqnarray}\label{ex_x1}
 t_2&=&-\frac{E_\beta}{(E_\beta^2-1)^2},\;\;\;
 t_1\;=\;\frac{1-3E_\beta^2}{(E_\beta^2-1)^2},\;\;\;
 x_1=\frac{E_\beta-2 E_\beta^3}{(E_\beta^2-1)^2}-{\mbox{arctanh}}(E_\beta),
 \\\label{ex_v1}
 \partial_{t_1} E_\beta &=& 
 \frac{(E_\beta^2-1)^3}{2 E_\beta(1+3 E_\beta^2)},\;\;\;
 \partial_{t_2} x_1 =\frac{E_\beta}{2},\;\;\;\partial_{t_1} t_2 =\frac{1}{2 E_\beta}.
 \end{eqnarray}
 Equations (\ref{ex_x1})  describe  one-dimensional line  in 3-dimensional space.

The similar analysis of the break points manifolds for the functions $P_{\alpha\beta}$ shows that these manifolds coincide with ${\cal{E}}_\beta$, i.e. they are described by the same system
(\ref{first_req},\ref{br}) supplemented by  both eqs.(\ref{sol_P_red}). To obtain this result, we need eq.(\ref{sol_P_red}b) where $P_{\alpha\beta}$, $t_1$ and $t_2$ are taken as independent variable while $x_1$ and  $E_\beta$ are considered as functions of them.

 \paragraph{Break points manifold for wave (\ref{sol_P}).}
 Consider the  wave profile breaking for the solutions of the eq.(\ref{sol_P}a) with time  $t_1$ and space variables $(x_1,x_2,t_2)$.
As usual, the first requirement to the break points  manifold ${\cal{E}}_\beta$ is  that the derivative of $E_\beta$  tends to infinity in some direction(s) in space $(x_1,x_2,t_2)$. Let us write all partial derivatives:
 \begin{eqnarray}\label{partial_3}
 \partial_{x_1}{E_\beta}&=&\frac{\partial_{z_1} F^{(E)}_\beta(z_1,z_2)|_{{z_1=x_1-t_1 E_\beta}\atop{
  z_2=x_2-t_2 E_\beta}}}{d}
  ,\;\;\;
  \partial_{x_2}{E_\beta}\;=\;\frac{\partial_{z_2} F^{(E)}_\beta(z_1,z_2)|_{{z_1=x_1-t_1 E_\beta}\atop{
  z_2=x_2-t_2 E_\beta}}}{d},\\\nonumber
  \partial_{t_1}{E_\beta}&=&-\frac{E_\beta \partial_{z_1} F^{(E)}_\beta(z_1,z_2)|_{{z_1=x_1-t_1 E_\beta}\atop{
  z_2=x_2-t_2 E_\beta}}}{d},\;\;\; \partial_{t_2}{E_\beta}=-\frac{E_\beta \partial_{z_2} F^{(E)}_\beta(z_1,z_2)|_{{z_1=x_1-t_1 E_\beta}\atop{
  z_2=x_2-t_2 E_\beta}}}{d}
 \\\nonumber
 &&
d= \left.\Big(1+t_1\partial_{z_1}F^{(E)}_\beta(z_1,z_2)+  t_2 \partial_{z_2}F^{(E)}_\beta(z_1,z_2) \Big)\right|_{{z_1=x_1-t_1 E_\beta}\atop{
  z_2=x_2-t_2 E_\beta}}.
 \end{eqnarray}
  We see that all derivatives tend to infinity if $d=0$:
 \begin{eqnarray}\label{Fzz1}
 \left.\Big(1+t_1\partial_{z_1}F^{(E)}_\beta(z_1,z_2)+  t_2 \partial_{z_2}F^{(E)}_\beta(z_1,z_2) \Big)\right|_{{z_1=x_1-t_1 E_\beta}\atop{
  z_2=x_2-t_2 E_\beta}}=0
 \end{eqnarray}
 Note that the derivative of $E_\beta$ is zero in the direction  $\vec t = \frac{1}{\sqrt{1+E_\beta^2}}(0,E_\beta,1)$. The second requirement is that function must change concavity in the break points. To relate this requirement with the second derivative of some coordinate with respect to $E_\beta$  we follow the usual strategy.
 Let  $E_\beta$ be independent variable. There are three   
 following cases:
 \begin{eqnarray}
\label{case_P2}
1. &&
  {\mbox{ $x_2$ is a function of $E_\beta$, $x_1$, $t_2$ and $t_1$,}}
  \\\label{case_P3}
2. &&
{\mbox{ $x_1$ is a function of $E_\beta$, $x_2$, $t_2$ and $t_1$,}}
  \\\label{case_P1}
3. &&
 {\mbox{ 
   $t_2$ is a function of $E_\beta$, $x_1$, $x_2$ and $t_1$}}.  
 \end{eqnarray}
Let us consider the first case in details. 
 One can find the first  derivatives of $x_2$ differentiating eq.(\ref{sol_P}) with respect to $E_\beta$, $x_1$ and $t_2$ (we do not need the derivative with respect to time $t_1$):
 \begin{eqnarray}\label{Ex}
 &&
 \partial_{E_\beta} x_2 =\left. \frac{1+ t_1\partial_{z_1}F^{(E)}_\beta(z_1,z_2)+t_2 \partial_{z_2}F^{(E)}_\beta(z_1,z_2)}{\partial_{z_2}F^{(E)}_\beta(z_1,z_2)}\right|_{{z_1=x_1-t_1 E_\beta}\atop{
  z_2=x_2-t_2 E_\beta}} ,\\\nonumber
 &&
 \partial_{x_1} x_2 =-\left. \frac{\partial_{z_1}F^{(E)}_\beta(z_1,z_2)}{\partial_{z_2}F^{(E)}_\beta(z_1,z_2)}\right|_{{z_1=x_1-t_1 E_\beta}\atop{
  z_2=x_2-t_2 E_\beta}} ,\;\;\;
 \partial_{t_2} x_2=E_\beta. 
 \end{eqnarray} 
The  requirement (\ref{Fzz1})  
 yelds  $\partial_{E_\beta} x_2=0$.
Similar to the previous paragraph, we take  the condition  $\partial_{E_\beta}^2 x_2=0$ as the second requirement defining  the   break points manifold. The explicit form of this requirement  may be found differentiating the eq.(\ref{Ex}a) with respect to $E_\beta$ and using eqs.(\ref{Ex}) for the first derivatives of $x_2$.  One gets in result:
\begin{eqnarray}\label{Fzz20}
&&
\left. \Big[ t_1 \partial_{z_2} F^{(E)}_\beta(z_1,z_2)\Big(
2 \big(1+t_1 \partial_{z_1} F^{(E)}_\beta(z_1,z_2)\big)  
\partial_{z_1}\partial_{z_2} F^{(E)}_\beta(z_1,z_2)-\right.\\\nonumber
&&
\left.
t_1 
\partial_{z_1}^2 F^{(E)}_\beta(z_1,z_2)\partial_{z_2} F^{(E)}_\beta(z_1,z_2)\Big)
-
\Big(1+t_1 \partial_{z_1} F^{(E)}_\beta(z_1,z_2)\Big)^2  
\partial_{z_2}^2 F^{(E)}_\beta(z_1,z_2)
\Big]
\right|_{{z_1=x_1-t_1 E_\beta}\atop{
  z_2=x_2-t_2 E_\beta}} =0.
  \end{eqnarray}
  This equation may be given a simpler form using (\ref{Fzz1}) in order to eliminate 
 $ \partial_{z_2} F^{(E)}_\beta(z_1,z_2)$ (if $t_2\neq 0$) or $ \partial_{z_1} F^{(E)}_\beta(z_1,z_2)$ (if $t_1\neq 0$):
\begin{eqnarray}\label{Fzz2}
\left. \Big(
t_1^2 \partial_{z_1}^2 F^{(E)}_\beta(z_1,z_2) +
 2 t_1 t_2 \partial_{z_1}\partial_{z_2} F^{(E)}_\beta(z_1,z_2) + 
t_2^2  \partial_{z_2}^2 F^{(E)}_\beta(z_1,z_2)\Big) 
\right|_{{z_1=x_1-t_1 E_\beta}\atop{
  z_2=x_2-t_2 E_\beta}} =0.
\end{eqnarray}
The equations (\ref{Fzz1},\ref{Fzz2}) must be supplemented by the eq. (\ref{sol_P}a).

Remark, that starting with the eq.(\ref{case_P3}) or eq.(\ref{case_P1}) we will end up with the same two conditions defining the break points manifold, eqs.(\ref{Fzz1},\ref{Fzz2}).

System (\ref{sol_P}a,\ref{Fzz1},\ref{Fzz2}) allows one to write the linear system defining the velocity of  the  break points manifold. 
Assuming that variables $E_\beta$, $x_1$, $x_2$ and $t_2$ depend on $t_1$ and differentiating eqs.(\ref{Fzz1},\ref{Fzz2})  with respect to $t_1$ we get:
\begin{eqnarray}\label{XTE_vel}
&&(E_\beta -x_1') \partial_{z_1}F^{(E)}_\beta(z_1,z_2)+
(E_\beta t_2'-x_2') \partial_{z_2}F^{(E)}_\beta(z_1,z_2)
=0,\\\nonumber
&&
\partial_{z_1}F^{(E)}_\beta(z_1,z_2)-E_\beta t_2 \partial_{z_1}\partial_{z_2} F^{(E)}_\beta(z_1,z_2)-E_\beta t_1 \partial_{z_1}^2 F^{(E)}_\beta(z_1,z_2)+
\\\nonumber
&&
x_1' 
\Big(t_2 \partial_{z_1}\partial_{z_2} F^{(E)}_\beta(z_1,z_2)+t_1 \partial_{z_1}^2 F^{(E)}_\beta(z_1,z_2)\Big)+
\\\nonumber
&&
x_2' \Big(t_2 
\partial_{z_2}^2 F^{(E)}_\beta(z_1,z_2)+t_1 \partial_{z_1}\partial_{z_2} F^{(E)}_\beta(z_1,z_2)\Big)+\\\nonumber
&&
t_2' \Big(\partial_{z_2}F^{(E)}_\beta(z_1,z_2)-E_\beta t_2 
\partial_{z_2}^2 F^{(E)}_\beta(z_1,z_2)-E_\beta t_1 \partial_{z_1}\partial_{z_2} F^{(E)}_\beta(z_1,z_2)\Big)
=0,
\\\nonumber
&&-E_\beta t_1^2  \partial_{z_1}^3 F^{(E)}_\beta(z_1,z_2) +2 t_1\partial_{z_1}^2 F^{(E)}_\beta(z_1,z_2) -2 E_\beta t_1t_2 \partial_{z_1}^2\partial_{z_2} F^{(E)}_\beta(z_1,z_2) 
+
\\\nonumber
&&2 t_2 \partial_{z_1}\partial_{z_2} F^{(E)}_\beta(z_1,z_2)-E_\beta t_2^2 \partial_{z_1}\partial_{z_2}^2 F^{(E)}_\beta(z_1,z_2)+
\\\nonumber
&&
x_1' \Big( t_1^2\partial_{z_1}^3 F^{(E)}_\beta(z_1,z_2)+2 t_1
t_2 \partial_{z_1}^2\partial_{z_2} F^{(E)}_\beta(z_1,z_2) +t_2^2 \partial_{z_1}\partial_{z_2}^2 F^{(E)}_\beta(z_1,z_2)\Big)+
\\\nonumber
&&
x_2' 
\Big(t_1^2\partial_{z_1}^2\partial_{z_2} F^{(E)}_\beta(z_1,z_2) +2 t_1t_2 \partial_{z_1}\partial_{z_2}^2 F^{(E)}_\beta(z_1,z_2) +t_2^2 
\partial_{z_2}^3 F^{(E)}_\beta(z_1,z_2)\Big)+
\\\nonumber
&&
t_2' \Big(-E_\beta t_1^2\partial_{z_1}^2\partial_{z_2} F^{(E)}_\beta(z_1,z_2) +2t_1 \partial_{z_1}\partial_{z_2} F^{(E)}_\beta(z_1,z_2) 
-2 E_\beta t_1t_2 \partial_{z_1}\partial_{z_2}^2 F^{(E)}_\beta(z_1,z_2) +\\\nonumber
&&
2 t_2 \partial_{z_2}^2 F^{(E)}_\beta(z_1,z_2)-E_\beta 
t_2^2 \partial_{z_2}^3 F^{(E)}_\beta(z_1,z_2)\Big)-
E_\beta' 
\Big(t_1^3\partial_{z_1}^3 F^{(E)}_\beta(z_1,z_2) +
\\\nonumber
&&
3 t_1^2t_2 \partial_{z_1}^2\partial_{z_2} F^{(E)}_\beta(z_1,z_2) +
3t_1 t_2^2 
\partial_{z_1}\partial_{z_2}^2 F^{(E)}_\beta(z_1,z_2) +t_2^3 \partial_{z_2}^3 F^{(E)}_\beta(z_1,z_2)\Big)=0
\end{eqnarray}
where 
$z_1=x_1-E_\beta t_1$, $z_2=x_2-E_\beta t_2$. 
 As a simple example, let 
 \begin{eqnarray}
 F^{(E)}_\beta(z_1,z_2)=-\tanh(z_1+z_2).
 \end{eqnarray}
 Then eqs.(\ref{sol_P}a,\ref{Fzz1},\ref{Fzz2},\ref{XTE_vel}) yield:
 \begin{eqnarray}\label{ex_x}
&&
 x_1=-x_2,\;\;t_2=1-t_1,\;\;E_\beta=0, \\\label{ex_v}
 &&
 x_1'=-x_2',\;\;\; t_2'=-1,\;\;\;E_\beta'=0.
 \end{eqnarray}
 Equations (\ref{ex_x}) describe  two-dimensional surface
 in 4-dimensional space; breaking starts
  at fixed value of  $E_\beta$:  $E_\beta=0$. 
 Similar analysis shows that ${\cal{E}}_\beta$ is  break points manifold corresponding to the functions $P_{\alpha\beta}$ (see eq.(\ref{sol_P}b)) as well.
 
\section{New version of the dressing method}
\label{Dressing}
  We have  described some solutions manifolds to well known $S$-integrable models (which were called PDE1($t_1,t_2;w^{(0)}$) in Sec.\ref{Algebraic}.  It is remarkable, that these solutions manifolds cover  full solutions spaces to the assotiated families of the matrix first-order  quasilinear PDEs integrable by the method of characteristics (which were called PDE0($t_m;n_0,w$), $m=1,2$). 

The purpose of this section is the construction of such dressing algorithm  which ($a$) would reproduce the solutions manifold to PDE1s available  by the classical dressing method and ($b$) would describe
the different solution  manifold  to  PDE1s in  spirit of the method of characteristics, i.e.
 we will derive non-differential equations implicitly 
describing  solutions to PDE1($t_1,t_2;w^{(0)}$) which are  solutions  to appropriate PDE0($t_m;n_0,w$), $m=1,2$, as well. Our consideration is based on the new version of the dressing method, which is subsequent development of the basic ideas  of 
\cite{SZ1,ZS2,ZS,Z2,Z3}. As a particular result, 
eqs.(\ref{w_C},\ref{w_C_G_P},\ref{w_C_G_Pred}) will be obtained.

Let us describe the basic objects appearing in the dressing algorithm.
In \cite{ZS2}, eq.(\ref{equ-w}) with $\rho=0$ has been derived  as a simplest example of the implementation of the dressing method based on the following   integral homogeneous equation
\begin{eqnarray}\label{u0}
\int\limits_D \Psi(\lambda,\nu;x) U(\nu;x) d\Omega(\nu)\equiv\Psi(\lambda,\nu;x)*U(\nu;x) =0,
\end{eqnarray}
where $\Psi$ and $U$ are    $Q\times Q$ matrix dressing and spectral  functions respectively,  $\lambda$ and $\nu$ are complex spectral parameters, $\Omega$ is some measure on the complex plane $\nu$, ''$*$'' means integration over $\nu$ and $x=(x_1,x_2,\dots,t_1,t_2\dots)$ is a set of independent variables of nonlinear PDEs. 

However, integral equation (\ref{u0}) is not appropriate for our purposes.
In order to derive PDE1s together with assotiated PDE0s, we shell  replace  the  equation (\ref{u0}) with the following integral homogeneous equation
\begin{eqnarray}\label{u1}
\Psi(\lambda,\nu;x)*U(\nu,\mu;x)=0,
\end{eqnarray}
where $(M+1)Q\times Q$ matrix spectral function $U$ depends on two spectral parameters; $\Psi$ is $Q\times (M+1)Q$ dressing function and kernel of the integral operator. Here integer parameter $M=\dim{\mbox{ker}}\;\Psi$.
Following the strategy of \cite{ZS2}, we assume that the solution of (\ref{u1}) is not unique but may be represented in the form:
\begin{eqnarray}\label{U_sol}
U(\lambda,\mu;x)=\sum_{j=1}^M U^{(h;j)}(\lambda,\nu;x)* f^{(j)}(\nu,\mu;x),
\end{eqnarray}
where $f^{(j)}(\nu,\mu;x)$, $j=1,\dots,M$, are arbitrary $Q\times Q$ matrix  functions of arguments and $U^{(h;j)}$, $j=1,\dots,M$,  are $M$ nontrivial linearly independent solutions of the homogeneous equation (\ref{u1}), i.e
\begin{eqnarray}
\sum_{j=1}^M U^{(h;j)}(\lambda,\nu;x)* S^{(j)}(\nu,\mu;x)=0 \;\;\;\Rightarrow
\;\;\; S^{(j)}(\nu,\mu;x)\equiv 0,\;\;j=1,\dots,M.
\end{eqnarray}
This assumption causes the single linear  relation among any $M+1$ independent solutions  $U^{(j)}$, $j=0,\dots,M$, of eq.(\ref{u1}):
\begin{eqnarray}\label{linU^i}
U^{(0)}(\lambda,\mu;x)=\sum_{j=1}^M U^{(j)}(\lambda,\nu;x)*F^{(j)}(\nu,\mu;x),
\end{eqnarray}
where  $F^{(j)}$ are some $Q\times Q$ matrix  functions. 
As we shell see, all $U^{(j)}$ are expressed in terms of the single solution  $U$  through some linear  operators $L^{(j)}$, either differential or non-differential: 
\begin{eqnarray}
U^{(j)}(\lambda,\mu;x)=L^{(j)}(\lambda,\nu)*U(\nu,\mu;x),\;\;j=0,\dots,M.
\end{eqnarray}
 Thus, eq.(\ref{linU^i}) represents a linear equation for the spectral function $U$. Besides, we will show that  $F^{(j)}$ may be expressed in terms of $U$  using the external $Q\times (M+1)Q$  dressing matrix function $G(\lambda,\mu;x)$, similar to \cite{ZS2}.

To increase dimensionality of the dressing functions (i.e., the number of variables $x_j$ which may be introduced arbitrarily in the dressing functions) and, consequently, to increase the dimensionality of the assotiated nonlinear PDEs, we 
 introduce the  $Q\times Q$ matrix function  ${\cal{A}}(\lambda,\mu)$ and $(M+1)Q\times (M+1)Q$ matrix function  $A(\lambda,\mu)$ 
 by the following generalized commutation relation involving $\Psi$:
\begin{eqnarray}\label{Acom} {\cal{A}}(\lambda,\nu)*\Psi(\nu,\mu;x)=\Psi(\lambda,\nu;x)*A(\nu,\mu).
 \end{eqnarray}
Similar to eq.(\ref{AAA}) we  define operators ${\cal{A}}^j$ and $A^j$ as follows: ${\cal{A}}^{j}=\underbrace{{\cal{A}}*\cdots*{\cal{A}}}_j$, $A^{j}=\underbrace{A*\cdots*A}_j$. Consequently,  functions ${\cal{A}}(\lambda,\mu)$ and $A(\lambda,\mu)$
 generate the following set of functions
 \begin{eqnarray}\label{Arho}
 {\cal{A}}^{(mj)}=\rho^{(mj)}({\cal{A}}),\;\;
 {{A}}^{(mj)}=\rho^{(mj)}({A}):\;\;
 \rho^{(mj)}({\cal{A}})*\Psi=\Psi*\rho^{(mj)}({A}) ,
 \end{eqnarray}
 where scalar functions $\rho^{(mj)}(z)$ are representable by a
 positive power series of $z$, see eq.(\ref{ser_rho}),
  so that ${\cal{A}}^{(mj)}$ and ${{A}}^{(mj)}$ are well defined $Q\times Q$ and $(M+1)Q \times (M+1)Q$  matrix functions respectively. 
 
 As usual, $x$-dependence of the spectral function $U$ appears through the $x$-dependence of the dressing functions. 
 Let $x$-dependence of $\Psi$ be given by the equation
 \begin{eqnarray}\label{x} \Psi_{t_m}(\lambda,\mu;x)+\sum_{j=1}^N{\cal{A}}^{(mj)}(\lambda,\nu)*\Psi_{x_j}(\nu,\mu;x)+{\cal{C}}^{(m)}(\lambda,\nu)*\Psi(\nu,\mu;x) =\\\nonumber
\Psi(\lambda,\nu;x)*C^{(m)}(\nu,\mu) ,
 \end{eqnarray}
where ${\cal{A}}^{(mj)}$, $A^{(mj)}$  do not depend on $x$; ${\cal{C}}^{(m)}$ and $C^{(m)}$ are $Q\times Q$ and $(M+1)Q \times (M+1)Q$  matrix functions respectively, ${\cal{A}}^{(m)}*{\cal{C}}-{\cal{C}}^{(m)}*{\cal{A}}\neq 0$, 
$A*C-C*A\neq 0$.
Equations (\ref{Acom}) and (\ref{x}) represent the overdetermined linear system for $\Psi$. Compatibility condition of (\ref{Acom}) and (\ref{x}) yelds: 
\begin{eqnarray}\label{PsiP}
&&
{\cal{P}}^{(m)} *\Psi = \Psi * P^{(m)},
\\\label{PsiP2}
&&
{\cal{P}}^{(m)}={\cal{A}}*{\cal{C}}^{(m)}-{\cal{C}}^{(m)}*{\cal{A}},
\;\;P^{(m)}=A*C^{(m)}-C^{(m)}*A.
\end{eqnarray}
In addition, equations (\ref{x}) with different values of $m$ are  compatible only if ${\cal{P}}^{(m)}=0$ and  $P^{(m)}=0$. In other words, if ${\cal{P}}^{(m)}\neq 0$ and  $P^{(m)}\neq 0$, then  derived nonlinear PDEs will not possess commuting flows, at least among PDE0s. 

Now we may obtain the set of different solutions to the homogeneous  eq. (\ref{u1}) applying 
${\cal{A}}^{m}*$ and 
$(\partial_{t_m} + \sum_{j=1}^N {\cal{A}}^{(mj)}*\partial_{x_j}+{\cal{C}}^{(m)}*)$ to (\ref{u1}).
One gets 
\begin{eqnarray}\label{EE}
\Psi(\lambda,\nu;x)*E^{(j;m)}(\nu,\mu;x)=0,\;\;j=1,2,
\end{eqnarray}
where
\begin{eqnarray}\label{H2}
&&
E^{(1;m)}(\lambda,\mu;x)=A^{m}(\lambda,\nu)*
U(\nu,\mu;x),\\\nonumber
&&\label{H3}
E^{(2;m)}(\lambda,\mu;x)=U_{t_m}(\lambda,\mu;x) + \sum_{j=1}^N A^{(mj)}(\lambda,\nu) *U_{x_j}(\nu,\mu;x) +C^{(m)}(\lambda,\nu)*U(\nu,\mu;x).
\end{eqnarray}
 Thus, $E^{(j;m)}$ are  desirable solutions of the integral homogeneous  equation.
 The  derivation  of  nonlinear PDEs significantly depends  on the
 value of the parameter $M$.
 Hereafter we consider  $M=1$.
 
\subsection{Simplest degeneration of the ker $\Psi$: $M=1$}
\label{Section:M_1}
In this case eq.(\ref{U_sol}) reads:
\begin{eqnarray}\label{U_sol1}
U(\lambda,\mu;x)= U^{(h;1)}(\lambda,\nu;x)* f^{(1)}(\nu,\mu;x).
\end{eqnarray}
It follows from the above discussion that we have to introduce so-called external dressing $Q\times 2Q$ matrix function $G(\lambda,\mu;x)$, whose prescription will be explored in  Sec.\ref{Dressing:PDE1}.
Let $G$ be defined  
by the next overdetermined system of linear equations:
\begin{eqnarray}\label{G_com}
 &&
G(\lambda,\nu;x)*A(\nu,\mu) = \hat A(\lambda,\nu)*G(\nu,\mu;x)+ 
 H_1(\lambda;x) H_2(\mu;x)
 ,\\\label{G}\label{G_x}
&&
 G_{t_m}(\lambda,\mu;x) +\sum_{j=1}^N G_{x_j}(\lambda,\nu;x)*A^{(mj)}(\nu,\mu) -G(\lambda,\nu;x)*C^{(m)}(\nu,\mu)=
 \\\nonumber
&&
 -\hat C^{(m)}(\lambda,\nu)*G(\nu,\mu;x) - \sum_{j=1}^{\tilde N} T^{(j)} G(\lambda,\nu;x)*\tilde A^{(mj)}(\nu,\mu),
 \end{eqnarray}
 where $\tilde N$ is some integer,  $\hat A$ and $H_1$ are $Q\times Q$, while $H_2$ is $Q\times 2 Q$  matrix functions;
 functions $\tilde A^{(mj)}=\tilde \rho^{(mj)}(A)$ are defined by the series (\ref{ser_trho}).
Require also
\begin{eqnarray}
T^{(m)}A=AT^{(m)},\;\;T^{(m)} H_1=H_1T^{(m)}.
\end{eqnarray}
Functions $H_1(\lambda;x)$ and $H_2(\mu;x)$ are called external dressing functions as well as function $G(\lambda,\mu;x)$.  
The  compatibility condition of eqs.(\ref{G_com}) and (\ref{G_x}) yelds:
\begin{eqnarray}\label{comp00}\label{comp02}\label{G_com2}
&& G(\lambda,\nu;x)*P^{(m)}(\nu,\mu)-\hat P^{(m)}(\lambda,\nu)*G(\nu,\mu;x)
 =\\\nonumber
&&-L^{(m)}_1 H_1(\lambda;x) H_2(\mu;x)-
\sum_{j=1}^N {H_1}_{x_j}(\lambda;x) H_2(\nu;x) *A^{(mj)}(\nu,\mu)-
H_1(\lambda;x) L^{(m)}_2 H_2(\mu;x) ,
\end{eqnarray}
where
\begin{eqnarray}
\label{PsiP3}
&&
\hat P^{(m)}=\hat A* \hat C^{(m)}-\hat C^{(m)}  *\hat A,\\\nonumber
&&
L^{(m)}_1 H_1(\lambda;x)={H_1}_{t_m }(\lambda;x) +\hat C^{(m)}(\lambda,\nu)*H_1(\nu;x),
\\\nonumber
&&
L^{(m)}_2 H_2(\mu;x)=
{H_2}_{t_m}(\mu;x)+\sum_{j=1}^N{H_2}_{x_j}(\nu;x) *A^{(mj)}(\nu,\mu) +\\\nonumber
&&
\sum_{j=1}^{\tilde N}T^{(j)} H_2(\nu;x) *\tilde A^{(mj)}(\nu,\mu) - H_2(\nu;x) *C^{(m)}(\nu,\mu).
\end{eqnarray}
In order to derive the eq.(\ref{comp00}), we differentiate the eq.(\ref{G_com}) with respect to $t_m$ and use eq. (\ref{G_x}) to eliminate $G_{t_m}$ and eq.(\ref{G_com}) to simplify the result. After all transformations we end up with eq.(\ref{comp00}).

The compatibility condition (\ref{comp00}) produces  equations defining $H_i$, $i=1,2$, in the following way.  Eq.(\ref{comp00}) must be identical in $\lambda$ and $\mu$.
Solving it, we must take into account that two spectral parameters are not separated in the LHS of the eq.(\ref{comp00}), while each term in the RHS of this equation separates two spectral  parameters.
 We suggest two different solution of the eq. (\ref{G_com2}):

1. Let
\begin{subequations}\label{case01}
\begin{eqnarray}\label{CH}\label{case01a}
&&
\hat A*H_1=0,\\\label{case01b}
&&
P^{(m)}=\rho^{(m0)}( A),\;\;
{\cal{P}}^{(m)}=\rho^{(m0)}({\cal{ A}}),\;\;
 \hat P^{(m)}=\rho^{(m0)}(\hat A)
,\\\label{H12}
&&
{H_1}_{t_m} +\hat C^{(m)}*H_1=0,\;\;\; {H_1}_{x_j}=0,\\\label{H22}
&&
{H_2}_{t_m}+\sum_{j=1}^N{H_2}_{x_j} *A^{(mj)} +\sum_{j=1}^{\tilde N}T^{(j)} H_2 *\tilde A^{(mj)}- H_2 *C^{(m)} =\\\nonumber
&&-H_2* \rho^{(m0)}(A)*A^{-1},
\end{eqnarray}
where functions  $\rho^{(m0)}$ are defined by the eq.(\ref{ser_rho}). 
\end{subequations}
Then eq.(\ref{comp02}) becomes:
\begin{eqnarray}\label{GCA}
G*\rho^{(m0)}(A) - \rho^{(m0)}(\hat A)*G= H_1 H_2 \rho^{(m0)}(A)*A^{-1},
\end{eqnarray}
which is equivalent to the eq.(\ref{G_com}) in virtue of the eq.(\ref{CH}). 
Compatibility of eqs.(\ref{G_x}) with different $m$ requires $P^{(m)}=0$ and $\hat P^{(m)}=0$.
These two conditions provide compatibility of  eqs.(\ref{case01}) with different $m$. Otherwise
$m$ takes a single value (say, $m=1$) and assotiated nonlinear PDEs have no commuting flows, at least among PDE0s.

 {\it Remark.} Eq.(\ref{GCA})
 requires that 
 positive series for $\rho^{(m0)}(z)$  start with $z^1$,
 which corresponds to $\alpha^{(m0;0)}=0$ in eq.(\ref{ser_rho}).
However, it will be shown in the Sec.(\ref{Section:matr}) (see a paragraph after the eq.(\ref{rho_w})),
that this is not a strong restriction. 
 
2. Let 
\begin{subequations}\label{case02}
\begin{eqnarray}\label{case02a}
&&
\hat A*H_1\neq 0,\;\;\;{\cal{C}}^{(m)}=\hat C^{(m)}=0,\;\;\;C^{(m)}=0,\\\label{H12_2}
&&
{H_1}_{t_m} =0,\;\;\; {H_1}_{x_j}=0,\;\;j=1,\dots,N,\\\label{H22_2}
&&
{H_2}_{t_m}+\sum_{j=1}^N{H_2}_{x_j} *A^{(mj)} +\sum_{j=1}^{\tilde N}T^{(j)} H_2 *\tilde A^{(mj)} =0.
\end{eqnarray}
\end{subequations}
Appropriate nonlinear  PDEs possess commuting flows since the eqs. (\ref{G_x}) with different $m$ are compatible, as well as  eqs.(\ref{case02}) with different  $m$.

It is evident that the eq.(\ref{PsiP}) is satisfied for both cases (\ref{case01}) and (\ref{case02}).

\subsubsection{Spectral system for $U(\lambda,\mu;x)$}
\label{Dressing:PDE1}
Now we are ready to derive the linear spectral system.
Eq. (\ref{U_sol1})   means that any two solutions of the homogeneous equation (\ref{u1}) 
are  linearly dependent. In particular, any solution $E^{(j;m)}$ 
(\ref{H2}) 
is   linearly dependent on $U$, i.e
\begin{eqnarray}\label{U_sp1}
&&
E^{(1;1)}(\lambda,\mu;x)=U(\lambda,\nu)*\tilde F(\nu,\mu;x),\;\; \Rightarrow \\\nonumber
&&
A(\lambda,\nu;x)*U(\nu,\mu;x) = U(\lambda,\nu;x)*\tilde F(\nu,\mu;x),
\\\label{U_sp2}
&&
E^{(2;m)}(\lambda,\mu;x)=U(\lambda,\nu)* F^{(m)}(\nu,\mu;x),\;\; \Rightarrow \\\nonumber
&&
U_{t_m}(\lambda,\mu;x) + \sum_{j=1}^N A^{(mj)}(\lambda,\nu) *U_{x_j}(\nu,\mu;x) +C^{(m)}(\lambda,\nu)*U(\nu,\mu;x)=\\\nonumber
&&
 U(\lambda,\nu;x)*F^{(m)}(\nu,\mu;x), \;\;\;m=1,2,\dots.
\end{eqnarray}
Eqs. (\ref{U_sp1},\ref{U_sp2}) represent the overdetermined linear system (the spectral system) for the spectral function $U(\lambda,\mu;x)$.
Remember that solution of eq.(\ref{u1}) is not unique. 
To obtain uniqueness we introduce   one more equation for the spectral function $U$. 
 This equation can be largely arbitrary, but, in order to derive
 the simplest nonlinear PDEs, we select the following 
 equation:
 \begin{eqnarray}\label{condition}\label{sol_GU}
G(\lambda,\nu;x)*U(\nu,\mu;x)=I\delta(\lambda-\mu),
\end{eqnarray}
so that $U$ is {\it the unique} solution of the system (\ref{u1},\ref{condition}). In other words, the equation (\ref{condition}) fixes the function $f^{(1)}(\lambda,\mu;x)$ in the eq.(\ref{U_sol1}).
Now, applying  $G*$ to the eqs.(\ref{U_sp1},\ref{U_sp2}), one gets the following expressions  for 
$\tilde F$ and $F^{(m)}$:
\begin{eqnarray}\label{F_def0}
\tilde F(\nu,\mu;x)&=&G(\lambda,\nu;x)*E^{(1;1)}(\nu,\mu;x)=\\\nonumber
&&\hat A(\lambda,\mu)+
H_1(\lambda;x)  H_2(\nu;x)*U(\nu,\mu;x),\\\nonumber
F^{(m)}(\nu,\mu;x)&=&
G(\lambda,\nu;x)*E^{(2;m)}(\nu,\mu;x)= 
\\\nonumber
&& \hat C^{(m)}(\lambda,\mu)+\sum_{j=1}^{\tilde N} T^{(j)} {{G}}(\lambda,\nu;x)*\tilde A^{(mj)}(\nu,\tilde \nu)*U(\tilde \nu,\mu;x)+ \\\nonumber
&&\sum_{j=1}^N \Big({G}(\lambda,\nu;x)*A^{(mj)}(\nu,\tilde \nu)*U(\tilde \nu,\mu;x)\Big)_{x_j},\;\;m=1,2,\dots .
\end{eqnarray}
Thus, eqs.(\ref{U_sp1},\ref{U_sp2}) become the overdetermined spectral system for the spectral function $U(\lambda,\mu;x)$. We consider the particular examples of the spectral systems and of the 
appropriate nonlinear PDEs  in Secs.\ref{Section:matr} and \ref{Section:matr_g}.

\paragraph{Spectral functions depending on single spectral parameter; fields and remarkable reductions.} 
\label{Section:M_1_sp}
Eqs.(\ref{U_sp1},\ref{U_sp2}) depend on two spectral parameters due to the spectral function $U(\lambda,\mu;x)$. However, functions of single spectral parameter  appear in the algorithm naturally. These functions are following:
\begin{eqnarray}
&&
V^{(j)}(\lambda;x)={U}(\lambda,\mu;x)*
\hat A^j(\mu,\nu)*H_1(\nu;x)
,\\\nonumber
&&
W^{(j)}(\mu;x)= H_2(\lambda;x)*A^{j}(\lambda,\nu)*U(\nu,\mu;x).
\end{eqnarray}
They satisfy the spectral equations with single spectral parameter which appear after applying $*\hat A^j*H_1$ and $H_2*A^j*$ to the eqs.(\ref{U_sp1},\ref{U_sp2}), see Secs.\ref{Section:matr},\ref{Section:matr_g}.

The dependent variables (or fields) of the nonlinear PDEs  are expressed in terms of the spectral and dressing functions by the next formulae:
\begin{eqnarray}\label{ww}
w^{(kn)}(x)=H_2(\lambda;x)*A^k(\lambda,\mu)*V^{(n)}(\mu;x)=
W^{(k)}(\mu;x)*\hat A^n(\mu,\nu)*H_1(\nu;x).
\end{eqnarray}
This definition of the fields suggests us the following two types of reductions:
\begin{eqnarray}\label{red22}
1. && 
\hat A^{n_0}*H_1=\sum_{j=1}^{n_0-1}\hat A^{j}*H_1r^{(j)} \;\;\;\Rightarrow
\\\nonumber
&&
w^{(kn_0)}=\sum_{j=1}^{n_0-1}w^{(kj)}r^{(j)},\;\;
\forall k,\;\;\;
\stackrel{k=0,(\ref{w_w})}{\Rightarrow}\;\;\;(\ref{red2}),
\end{eqnarray}
\begin{eqnarray}\label{red11}\label{redH2_1}
2. && H_2*A^{k_0}= \sum_{j=1}^{k_0-1} r^{(j)} H_2*A^{j} \;\;\;\Rightarrow
\\\nonumber
&&
w^{(k_0n)}= \sum_{j=1}^{k_0-1} r^{(j)} w^{(jn)},\;\;\forall n\;\;\;\stackrel{n=0,(\ref{tilde_w_w})}{\Rightarrow}\;\;\;(\ref{red1}).
\end{eqnarray}
It will be shown in Sec.(\ref{sol_PDE0}) (see text after the eq.(\ref{R_def})), that  $r^{(i)}$ must be scalar constants (remember that  these parameters  are matrices in the algebraic approach, see the paragraph after the eqs.(\ref{wwww})).
We consider only the reduction (\ref{red22}).

In accordance with Introduction,  we refer to the nonlinear PDEs corresponding to the reduction (\ref{red22}) as PDE0s, while the  nonlinear PDEs without any reduction will be referred to as PDE1s.
Although the nonlinear PDEs may be derived for any given functions $\rho^{(mj)}$ and $\tilde\rho^{(mj)}$, it is difficult to write the nonlinear PDEs keeping these functions unfixed, in general.
The  example considered in Sec.\ref{Section:matr} is an 
exception. 

\subsubsection{Case (\ref{case01}).
Quasilinear first order matrix equations solvable by
the method of characteristics. }
\label{Section:matr}
In this section, we briefly  reproduce results obtained in \cite{SZ1,ZS2} by a different method. 
First of all, remark that the eq.(\ref{case01}a) is nothing but the reduction (\ref{red22}) with $n_0=1$. Thus, the derived equations will be  PDE0s. 
In this case eqs.(\ref{U_sp1},\ref{U_sp2}) yield:
\begin{eqnarray}\label{U_sp1_case01}
&&
A*U=U*\hat A+U*H_1 H_2*U ,\\\nonumber
&&
U_{t_m}+\sum_{j=1}^N A^{(mj)}* U_{x_j}+C^{(m)}*U = U*\Big[
\hat C^{(m)} +\\\nonumber
&&
\sum_{j=1}^{\tilde N}T^{(j)}\Big( \tilde A^{(mj)} +  H_1 H_2*\tilde A^{(mj)}*A^{-1}*U\Big)+\sum_{j=1}^{ N}
(H_1 H_2*A^{(mj)}*A^{-1}*U)_{x_j}\Big)\Big].    
\end{eqnarray}
 Applying $*H_1$ to the eqs.(\ref{U_sp1_case01}) one gets
 ($V=V^{(0)}$, $w=w^{(00)}$)
 \begin{eqnarray}\label{U_sp1_case01_2}
&&
A*V=V w \;\;\Rightarrow \;\; A^{(mj)}*V=V \rho^{(mj)}(w),
\\\label{linV0}
&&
V_{t_m}(\lambda;x) + \sum_{j=1}^N A^{(mj)}(\lambda,\nu)* V_{x_{j}}(\nu;x) +C^{(m)}(\lambda,\nu)*V(\nu;x) = \\\nonumber
&&
V(\lambda;x) \sum_{j=1}^{\tilde N} T^{(j)} \tilde \rho^{(mj)}(w)+V(\lambda;x) \sum_{j=1}^N \rho^{(mj)}_{x_{j}}(w).
\end{eqnarray}
or,  substituting the eq. (\ref{U_sp1_case01_2}) into the eq.(\ref{linV0}), one has:
\begin{eqnarray}\label{linV}
V_{t_m}(\lambda;x) + \sum_{j=1}^N V_{x_{j}}(\lambda;x) \rho^{(mj)}(w)+C^{(m)}(\lambda,\nu)*V(\nu;x) = \\\nonumber
V(\lambda;x) \sum_{j=1}^{\tilde N} T^{(j)} \tilde \rho^{(mj)}(w) .
\end{eqnarray}
After applying $H_2*$ to this equation one ends up 
with PDE0:
\begin{eqnarray}\label{rho_w}
w_{t_m} + \sum_{j=1}^N w_{x_{j}} \rho^{(mj)}(w) +\rho^{(m0)}(w)= [w, \sum_{j=1}^{\tilde N} T^{(j)} \tilde \rho^{(mj)}(w) ].
\end{eqnarray}
It was mentioned above, that this equation has commuting flows only if $\rho^{(m0)}=0$. Then $m=1,2,\dots$ in this equation. Otherwise, $m$ must be fixed, say, $m=1$. 

In accordance with  the {\it Remark} given after the eq.(\ref{GCA}), power series for $\rho^{(10)}(z)$ must start with $z^1$. Otherwise, multiplying eq.(\ref{rho_w}) by $w$ from the right and replacing $t_1\to x_{N+1}$ we result in the PDE0  having the form of ''stationary'' eq.(\ref{rho_w}) satisfying the condition of the above remark.

Eqs. (\ref{U_sp1_case01_2}) and (\ref{linV0}) has to be taken as the spectral system with the spectral function $V(\lambda;x)$ corresponding to the nonlinear PDE (\ref{rho_w}).
In \cite{SZ1} 
this equation has been derived with fixed $m=1$,  $\rho^{(1j)}\equiv \rho^{(j)}$, $j>0$,  $\rho^{(10)}\equiv \rho$,
$\tilde \rho^{(1j)}=\tilde \rho\delta_{1j}$. In addition,  $\rho=0$ in \cite{ZS2}, where another variant of the dressing method  has been introduced.
 However, both techniques developed 
in \cite{SZ1} and \cite{ZS2} may be applied to the eq.(\ref{rho_w}) 
as well.

\subsubsection{Case (\ref{case02}). PDE1s and assotiated PDE0s: 
$N$-wave and Pohlmeyer equations}
\label{Section:matr_g}\label{Char_Gen}
The main feature of the eqs. (\ref{case02}) is that they set 
$C^{(m)}=0$ and do not force the reduction (\ref{red22}). Nonlinear PDEs corresponding to this case  give rise to 
 PDE1s.
Then, been imposed, reduction (\ref{red22}) reduces PDE1s to  PDE0s.

 We study nonlinear PDEs corresponding to the  particular choice of the functions $\rho^{(mj)}$ and $\tilde\rho^{(mj)}$.
For instance, let $A^{(mj)}=s^{(m)} A\delta_{mj}$,  $\tilde A^{(mj)}=A\delta_{mj}$,   $s^{(m)}$ are scalar constants. 
Then the spectral system with two spectral parameters, eqs.(\ref{U_sp1},\ref{U_sp2}), reads:
\begin{eqnarray}
&&\label{U_sp30}
A*U=U*\hat A + V^{(0)} W^{(0)},\\\nonumber
&&
U_{t_m} + s^{(m)}A*U_{x_m} = 
U*\hat A T^{(m)}+
V^{(0)} T^{(m)} W^{(0)}+ s^{(m)} V^{(0)} W^{(0)}_{x_m},\;\;m=1,2.
\end{eqnarray}
The appropriate spectral system for the spectral functions $V^{(n)}(\lambda;x)$ with 
 single spectral parameter appears after   applying  
$*\hat A^n*H_1$, $n=0,1,\dots$, to the eqs.(\ref{U_sp30}):
\begin{eqnarray}\label{U_sp21}
&&
A(\lambda,\nu;x)*V^{(n)}(\nu;x) = V^{(n+1)}(\lambda;x) +
V^{(0)}(\lambda;x)  w^{(0n)}(x),
\\\nonumber
&&
V^{(n)}_{t_m}(\lambda;x) + s^{(m)} \Big(A(\lambda,\nu) *V^{(n)}_{x_m}(\nu;x) -
V^{(0)}(\lambda;x)  w^{(0n)}_{x_m}(x)\Big) =
V^{(n+1)}(\lambda;x) T^{(m)}+\\\nonumber
&&
V^{(0)}(\lambda;x) T^{(m)} w^{(0n)}(x),\;\;m=1,2.
\end{eqnarray}
In principle, applying $H_2*A^{k}*$ to the eq.(\ref{U_sp30}), one could get the linear equations for another set of  functions of single spectral parameter $W^{(k)}(\mu;x)$. However, these equations will not be used, so that we do not represent them here.

The 
 system of nonlinear PDEs  may be derived  
 applying $H_2*A^{k}*$ to the eq.(\ref{U_sp21}).
\begin{eqnarray}
\label{w0n+1_2}
&&
w^{((k+1)n)} =w^{(k(n+1))} + w^{(k0)} w^{(0n)},\\
\label{w0n+12}
&&w^{(kn)}_{t_m} + s^{(m)}(w^{((k+1)n)}_{x_m}-w^{(k0)} w^{(0n)}_{x_m}) =\\\nonumber
&&
 w^{(k(n+1))} T^{(m)}-T^{(m)} w^{((k+1)n)}  + w^{(k0)} T^{(m)} w^{(0n)},\;\;m=1,2.    
\end{eqnarray}
Using eq.(\ref{w0n+1_2}) to eliminate $w^{((k+1)n)}$
from the eq.(\ref{w0n+12})  and putting $k=0$ in the result we get (see eq.(\ref{w_w}) for definition of $w^{(n)}$):
\begin{eqnarray}\label{U_sp220_2}
&&
w^{(n)}_{t_m} + s^{(m)}(w^{(n+1)}_{x_m}+w^{(0)}_{x_m}
 w^{(n)})-[ w^{(n+1)},T^{(m)}] + 
  [T^{(m)},w^{(0)}]w^{(n)}=
 0,\\\nonumber
 &&
 n=0,1,\dots,\;\;m=1,2.
\end{eqnarray}
Fixing $n=0$ and eliminating $w^{(1)}$ from this system of two PDEs,
we get PDE1($t_1,t_2;w^{(0)}$).
Reduction (\ref{red22}) yields PDE0($t_m;n_0,w$), $m=1,2$:
\begin{eqnarray}\label{U_sp220_2red}
&&
w_{t_m} + s^{(m)}w_{x_m}
 w+ 
  [T^{(m)}_{n_0},w]w=
 0,\;\;m=1,2.
\end{eqnarray}

From another point of view, eliminating $w^{(k(n+1))}$ from the system (\ref{w0n+12}) and putting $n=0$ one gets:
\begin{eqnarray}\label{w0n+1_3}
&&
\tilde w^{(k)}_{t_m} +s^{(m)}( \tilde w^{(k+1)}_{x_m}-\tilde w^{(k)}\tilde w^{(0)}_{x_m})+
[T^{(m)},\tilde  w^{(k+1)}]
  +\tilde w^{(k)} 
  [\tilde w^{(0)},T^{(m)}] =
 0,\\\nonumber
 &&
 k=0,1,\dots,\;\;m=1,2,
\end{eqnarray}
where fields $\tilde w^{(n)}$ are  defined by the eq.(\ref{tilde_w_w}).
Fixing $k=0$ and eliminating $\tilde w^{(1)}$ from this system of two PDEs,
we get the same PDE1($t_1,t_2;w^{(0)}$).
Reduction (\ref{red11}) yields PDE0($t_m;k_0,\tilde w$), $m=1,2$:
\begin{eqnarray}\label{w0n+1_3red}
&&
\tilde w_{t_m} +s^{(m)} \tilde w\tilde w_{x_m}
  +\tilde w 
  [\tilde w,\tilde T^{(m)}_{k_0}] =
 0,
\;\;m=1,2.
\end{eqnarray}
$N$-wave and Pohlmeyer equations with apporpriate PDE0s and spectral problems follow from   eqs.(\ref{U_sp21}-\ref{w0n+1_3red}) with 
  $s^{(m)}=0$  (see Sec.\ref{Sec:Nw}) and  $T^{(m)}=0$, $s^{(m)}=1$
 (see Sec.\ref{Sec:P}) respectively.

Eq.(\ref{w0n+1_3}) can be formally obtained  by the 
transposition of the eq.(\ref{U_sp220_2}) with replacements 
$w^T \to -\tilde w$, $T^{(m)} \to - T^{(m)}$, similarly to 
the relations between eqs.(\ref{U_sp220_Nw_tilde}) and 
(\ref{U_sp220_Nw}) and between eqs.(\ref{U_sp220_alg_tilde}) and 
(\ref{U_sp220_alg}). Because of this similarity, 
 we will study only 
eqs.(\ref{U_sp220_2},\ref{U_sp220_2red}).


\subsection{Solutions space described by the dressing method}
\label{Solutions}
\label{sol_PDE1}
The dressing algorithm   describing solutions space to PDE1s 
consists of two basic parts:
\begin{enumerate}
\item\label{first}\label{1}
{\it The non-evolutionary part}, where we do not need the particular dependence of the dressing functions on $x_i$ and $t_i$. As a result, we derive the system of (non-differential) equations which  is valid for any possible $x$-dependence of the dressing functions. 
\begin{enumerate}
\item \label{1a} Solve the eq.(\ref{Acom}) as an equation defining $\Psi$.
\item \label{1b} \label{b} Solve the spectral equation (\ref{U_sp30}a) as an equation establishing the restrictions for the spectral function $U$.
\item \label{1c}  Solve the eq.(\ref{G_com}) as an equation for $G$.
\item \label{1d} Solve the eqs.(\ref{u1}) and (\ref{condition}) as equations for $U$ taking into account result of n.\ref{b}. 
\end{enumerate}
\item\label{2}
{\it The evolutionary part}, where we presume the special $x$-dependence  of the dressing functions.
\begin{enumerate}
\item\label{2a}
Solve eq. (\ref{x}) as an equation defining $x$-dependence of the internal dressing function $\Psi(\lambda,\mu;x)$.
\item\label{2b}
Solve one of the systems (\ref{case01}) or (\ref{case02}) defining $H_i$, $i=1,2$.
This step defines  $x$-dependence of the external dressing function $G(\lambda,\mu;x)$ owing to the eq.(\ref{G_com}). Eq. (\ref{G_x}) will be automatically satisfied.
\end{enumerate}
\end{enumerate} 
Details of the dressing algorithm producing the classical solutions manifold to PDE1s  are given in Sec.\ref{non-ev}.
In  order to describe solutions manifold to PDE1s  assotiated with reduction (\ref{red22}) 
  one has to evaluate the same algorithm with 
 eqs.(\ref{u1}, \ref{condition},\ref{U_sp30}a) modified by the reduction (\ref{red22}), see Sec.\ref{sol_PDE0} for details.

Before describing all steps of the algorithm in details, 
we give a description of the particular representation of the dressing and spectral functions which is necessary  hereafter. 
First of all, in order to fit the  requirement $M=1$, we 
introduce the next block-matrix representation of the functions:
\begin{eqnarray}\label{vec}
\Psi(\lambda,\mu;x)=[
\psi_{0}(\lambda,\mu;x) \;\;
\psi_{1}(\lambda,\mu;x)],&&
U(\lambda,\mu;x)=\left[\begin{array}{c}
u_{0}(\lambda,\mu;x) \cr
u_{1}(\lambda,\mu;x)\end{array}\right],\\\nonumber 
V^{(j)}(\lambda;x)=\left[\begin{array}{c}
v^{(j)}_{0}(\lambda;x)\cr 
v^{(j)}_{1}(\lambda;x)\end{array}\right],&&
W^{(j)}(\mu;x)=
w^{(j)}_0(\mu;x),\\\nonumber
G(\lambda,\mu;x)=[
g_{0}(\lambda,\mu;x)\;\; g_{1}(\lambda,\mu;x)],&&
H_2(\lambda;x)=[
h_{20}(\lambda;x) \;\;
h_{21}(\lambda;x)],
\\\nonumber
H_1(\lambda;x)= h_1(\lambda;x).&&
\end{eqnarray}
Any function in the RHS of formulae (\ref{vec}) is  $Q\times Q$ matrix function, so that $\Psi$, $G$ and $H_2$ are $Q\times 2 Q$ matrix functions; $U$ and $V^{(j)}$ are $2 Q\times Q$ matrix functions;
$W^{(j)}$ and $H_1$ are $Q\times Q$ matrix functions.
We also have to fix the  functions  ${\cal{A}}(\lambda,\mu)$, $A(\lambda,\mu)$ and $\hat A(\lambda,\mu)$  (compare with eq.(\ref{A_delta})):
\begin{eqnarray}\label{AAA0}
{\cal{A}}(\lambda,\mu) =\hat A(\lambda,\mu)= \lambda
\delta(\lambda-\mu)I
,\;\;\;\;
A(\lambda,\mu)=
\lambda \delta(\lambda-\mu)I_2.
\end{eqnarray}

\subsubsection{Dressing algorithm: classical solutions manifolds to PDE1s}
\label{non-ev}
\paragraph{Non-evolutionary part of the dressing algorithm.}
We describe all steps of the non-evolutionary part of the algorithm given in the beginning of the section (\ref{Solutions}) for the case (\ref{case02}) without reduction (\ref{red22}).

(\ref{1a}) Function   $\Psi$ is defined by  the eq.(\ref{Acom}), which  reads
\begin{eqnarray}\label{arb_sol_sp2}
(\lambda-\mu){\Psi}(\lambda,\mu;x)=
0 .
 \end{eqnarray}
 Its solution is following:
\begin{eqnarray}\label{hatPsi}
 \Psi(\lambda,\mu;x) =\hat \Psi(\lambda;x) \delta(\lambda-\mu),\;\;
\hat\Psi(\lambda;x)=[\hat \psi_{0}(\lambda;x)\;\; \hat \psi_{1}(\lambda;x)],
\end{eqnarray}
where $\hat \psi_{i}$, $i=0,1$, are $Q\times Q$ matrix functions.

(\ref{1b}) As it was  mentioned above, the main feature of the spectral system described in this paper is the presence of the spectral equation which has no derivatives with respect to $x$, see eq.(\ref{U_sp30}a):
\begin{eqnarray}
U(\lambda,\mu;x) (\lambda-\mu) =  V^{(0)}(\lambda;x) W^{(0)}(\mu;x).
\end{eqnarray}
This suggests us the next representation  for $U$:
\begin{eqnarray}\label{sol_U}
U(\lambda,\mu;x)=\frac{ V^{(0)}(\lambda;x) W^{(0)}(\mu;x)}{\lambda-\mu} + 
 U_0(\lambda;x)\delta(\lambda-\mu),
\;\;
U_0(\lambda;x)=\left[\begin{array}{c}u_{00}(\lambda;x)\cr
 u_{01}(\lambda;x)\end{array}\right],
\end{eqnarray}
where $u_{0i}$, $i=0,1$, are $Q\times Q$  matrix functions.

(\ref{1c}) Similarly, eq.(\ref{G_com}) reads 
\begin{eqnarray}\label{GHH}
&&
G(\lambda,\mu;x)(\lambda-\mu)= -H_1(\lambda;x)  H_2(\mu;x)
\end{eqnarray}
which suggest us the next formula for $G$:
\begin{eqnarray}\label{sol_G}
G(\lambda,\mu;x)=-\frac{H_1(\lambda;x) H_2(\mu;x)}{\lambda-\mu}+  G_0(\lambda;x)\delta(\lambda-\mu),\;\;G_0(\lambda;x)=[g_{00}(\lambda;x)\;\; g_{01}(\lambda;x)],
\end{eqnarray}
where $g_{0i}$, $i=0,1$, are $Q\times Q$  matrix functions and $G_0(\lambda;x)\delta(\lambda-\mu) $ is a solution of the eq.(\ref{G_x}).

(\ref{1d})
Next, substituting the eqs.(\ref{sol_U}) and (\ref{sol_G}) into the eq.(\ref{sol_GU}) 
one gets
\begin{eqnarray}\label{sol_HG0}
&&
-H_1(\lambda;x) \int\limits_D d\nu \frac{H_2(\nu;x)   V^{(0)}(\nu;x) W^{(0)}(\mu;x)}{(\lambda-\nu)(\nu-\mu)} -  H_1(\lambda;x) \frac{ H_2(\mu;x) U_0(\mu;x)}{\lambda-\mu}+\\\nonumber
&& \frac{G_0(\lambda;x)   V^{(0)}(\lambda;x)  W^{(0)}(\mu;x)}{\lambda-\mu} +
 G_0(\lambda;x) U_0(\lambda;x)\delta(\lambda-\mu) =
 I\delta(\lambda-\mu).
\end{eqnarray}
This equation must be identity for any $\lambda$ and $\mu$, which suggests us to split the eq.(\ref{sol_HG0}) into two  matrix  equations.
The first equation reads:
\begin{eqnarray}\label{sol_G_0}
&&
 G_0(\lambda;x)  U_0(\lambda;x) = I,
\end{eqnarray}
which is the system of $Q^2$ scalar equations for $2Q^2$ elements of $ U_0$, i.e. underdetermined  system of scalar equations.
The second equation reads:
\begin{eqnarray}\label{EEE}
&&
\frac{1}{\lambda-\mu}\left[ E_1(\lambda;x) W(\mu;x) +  H_1(\lambda;x) E_2(\mu;x)\right]=0,\\\nonumber
&&
E_1(\lambda;x)=G_0(\lambda;x) V^{(0)}(\lambda;x)-H_1(\lambda;x)\int\limits_D d\nu \frac{ 
 H_2(\nu;x)  V^{(0)}(\nu;x)}{\lambda-\nu}  - H_1(\lambda;x)\\\nonumber
&&
E_2(\mu;x)=-\int\limits_D d\nu\frac{
 H_2(\nu;x)  V^{(0)}(\nu;x)   W^{(0)}(\mu;x)}{\nu-\mu} -  H_2(\mu;x)   U_0(\mu;x) +  W^{(0)}(\mu;x)
\end{eqnarray}
In turn, the eq.(\ref{EEE}) must be splited into two matrix equations:
\begin{eqnarray}\label{E10}
&&
E_1(\lambda;x)=0,\\\label{E20}
&&
E_2(\mu;x)=0.
\end{eqnarray}
Due to our choice of the last term in the expression for $E_2$, 
eq.(\ref{E20}) coincides with eq.(\ref{sol_U}) after application $H_2*$ to it, which is necessary condition. The last term in expression for $E_1$ compensates the last term of $E_2$ after substitution in eq.(\ref{EEE}).

The matrix equation (\ref{E10}) represents $Q^2$ scalar equations for $2 Q^2$ elements of the matrix function $  V^{(0)}$, while the matrix equation (\ref{E20}) represents $Q^2$ scalar equations for $ Q^2$ elements of the matrix function $ W^{(0)}$. Thus, matrix equation (\ref{E10}) is underdetermined systems of scalar equations.  

The rest of equations for elements of $  V^{(0)}$ and $U_0$ follows from  the eq.(\ref{u1})  after substitution  the eq.(\ref{hatPsi}) for $ \Psi$ and  the eq.(\ref{sol_U}) for $ U$:
\begin{eqnarray}\label{PsiVW}
 \hat \Psi(\lambda;x) \frac{ V^{(0)}(\lambda;x) W^{(0)}(\mu;x)}{\lambda-\mu}+ \hat \Psi(\lambda;x)  U_0(\lambda;x) \delta(\lambda-\mu) =0. 
\end{eqnarray}
Since eq.(\ref{PsiVW}) must be identity for any $\lambda$ and $\mu$, it 
must be splitted
into  the following equations for $ V^{(0)}$ and $ U_0$:
\begin{eqnarray}\label{PsiV}
&&
\hat\Psi(\lambda;x) V^{(0)}(\lambda;x) =0,\\\label{PsiU0}
&&
\hat\Psi(\lambda;x) U_0(\lambda;x)=0
\end{eqnarray}
Each of these matrix equations represents $Q^2$ scalar equations 
for $2Q^2$ elements of the matrix functions  $ V^{(0)}$ and 
$U_0$ respectively. Thus, the system 
(\ref{sol_G_0},\ref{E10},\ref{E20},\ref{PsiV},\ref{PsiU0}) is 
the complete system for elements of $  V^{(0)}(\lambda;x)$, $
W^{(0)}(\mu;x)$ and $  U_0(\lambda;x)$. 
Having these functions,  the function $ U(\lambda,\mu;x)$ may be constructed using the formula (\ref{sol_U}). This ends the non-evolutionary part of the dressing algorithm.

All in all, in order to construct  the spectral function $U$, one has to solve 
\begin{eqnarray}\label{star}
1.&& {\mbox{
the eqs. (\ref{sol_G_0}) and (\ref{PsiU0}) for $ U_0$}},\\\nonumber
2.&& {\mbox{
the eqs. (\ref{E10}) and (\ref{PsiV}) for $ V^{(0)}$}},\\\nonumber
3.&& {\mbox{
the eq.(\ref{E20}) for $W^{(0)}$}}.
\end{eqnarray}
After the function $U$ has been constructed, one must use  the
definition (\ref{ww}) in order to construct the solution to PDE1($t_1,t_2;w^{(0)}$), i.e. the function $w^{(00)}\equiv w^{(0)}$.
However, it is simple to observe, that essentially important for construction of $w^{(0)}=H_2*V^{(0)}$ are eqs.(\ref{E10}) and (\ref{PsiV}) defining $V^{(0)}$.
Moreover, eq.(\ref{E10}) is equivalent to the classical $\bar\partial$-problem for Self-dual type $S$-integrable  equations. Let us write this equation in  standard form. To simplify derivation,   we take 
\begin{eqnarray}\label{class_H1}
H_1(\lambda;x)=I,
\end{eqnarray}
which is consistent with eq.(\ref{case02}b).
Eq.(\ref{PsiV}) yelds
\begin{eqnarray}
 v^{(0)}_1(\lambda;x)=-\hat\psi^{-1}_{1}(\lambda;x)\hat\psi_0(\lambda;x) v^{(0)}_0(\lambda;x).
\end{eqnarray}
Then eq.(\ref{E10}) gets the next form:
\begin{eqnarray}\label{inter_1}
&&
\phi(\lambda;x) v^{(0)}_0(\lambda;x)+\int\limits_D d\nu \frac{ 
\chi(\nu;x)  v^{(0)}_0(\nu;x)}{\lambda-\nu}  = I,
\end{eqnarray}
where
\begin{eqnarray}
&&
\phi(\lambda;x)=g_{00}(\lambda;x)-g_{01}(\lambda;x)
\hat\psi^{-1}_{1}(\lambda;x)\hat\psi_0(\lambda;x),
\\\nonumber
&&
\chi(\lambda;x)= -h_{20}(\lambda;x)+ h_{21}(\lambda;x)\hat\psi^{-1}_{1}(\lambda;x)\hat\psi_0(\lambda;x).
\end{eqnarray}
Introduce the new spectral function 
\begin{eqnarray}
v(\lambda;x)=\phi(\lambda;x)  v^{(0)}_0(\lambda;x).
\end{eqnarray}
Then eq.(\ref{inter_1}) yelds
\begin{eqnarray}\label{int_dbar}
\int\limits_D d\nu \hat R(\lambda,\nu;x) v(\nu;x) = I,
\end{eqnarray}
where
\begin{eqnarray}\label{dbar}
&&\hat R(\lambda,\nu;x) = \frac{R(\nu;x) }{\lambda-\nu} +I\delta(\lambda-\nu),\;\;\;R(\nu;x)=\chi(\nu;x)\phi^{-1}(\nu;x).
\end{eqnarray}
Here $R(\nu;x)$  is a new dressing function and $\hat R$ is a new kernel of the integral operator. By construction, $\dim{\mbox{ker}} \,\hat R =0$, so that eq.(\ref{int_dbar}) is uniquely solvable for $v$. 
Linear PDE for $R$ follows from the linear PDEs for  the functions  $\hat \psi_i$, $ g_{0i}$ and $h_{2i}$, $i=0,1$. These PDEs are 
eqs.(\ref{x}), (\ref{G_x}) and (\ref{case02}c), which read  (taking into account eq.(\ref{case02}a)):
\begin{eqnarray}\label{psi_diff0}
&&\Big({\hat \psi_i}(\lambda;x)\Big)_{t_m} +\sum_{j=1}^N \rho^{(mj)}(\lambda)
 \Big({\hat \psi_i} (\lambda;x) \Big)_{x_j}
 =0,\\\nonumber
\label{g0_diff0}
&&\Big({ g_{0i}}(\lambda;x)\Big)_{t_m} +\sum_{j=1}^N \rho^{(mj)}(\lambda)
\Big( { g_{0i}} (\lambda;x)\Big)_{x_j}
 +
 \sum_{j=1}^{\tilde N}T^{(j)} \tilde \rho^{(mj)}(\lambda)\; { {g_{0i}}}(\lambda;x) =0, 
\\\label{G_diff0}
&&
\Big({ {h}_{2i}}(\lambda;x)\Big)_{t_m} +
\sum_{j=1}^N \rho^{(mj)}(\lambda)\; \Big({{h}_{2i}}(\lambda;x)\Big)_{x_j} +
\sum_{j=1}^{\tilde N}T^{(j)} \tilde \rho^{(mj)}(\lambda)\; { {h}_{2i}}(\lambda;x)
=0,
\end{eqnarray}
where $i=0,1$.
From this system, we obtain the linear PDE for $R$:
\begin{eqnarray}
\label{Psi_diff0}
&&
R_{t_m}(\lambda;x) +
\sum_{j=1}^N \rho^{(mj)}(\lambda)\; R_{x_j}(\lambda;x) +
\sum_{j=1}^{\tilde N} \tilde \rho^{(mj)}(\lambda)\; [T^{(j)},R(\lambda;x)]
=0,
\end{eqnarray}
Eqs.(\ref{int_dbar},\ref{dbar},\ref{Psi_diff0}) represent the classical $\bar\partial$-problem for PDE1s,  \cite{ZM0,Z0}.
Eq.(\ref{ww}) for field $w^{(0)}$
reduces to the next one:
\begin{eqnarray}
w^{(0)}(x)=\int\limits_D d \lambda\;R(\lambda;x)v(\lambda;x) .
\end{eqnarray}

\paragraph{Evolutionary part of the dressing  algorithm.}
In this case, nn.\ref{2a},\ref{2b} of the dressing algorithm  reduce to the  solution of the single eq.(\ref{Psi_diff0}):
\begin{eqnarray}
 \label{dbar_sol_Psi}
&&
R(\lambda;x)=\\\nonumber
&&
\int\limits_{{\cal{D}}} dq\;
 e^{I\sum\limits_{j=1}^N \Big(q_j x_j - 
\sum\limits_{m=1}^2\rho^{(mj)}(\lambda) q_j t_m\Big) - 
\sum\limits_{j=1}^{\tilde N}\sum\limits_{m=1}^2 T^{(j)}  \tilde \rho^{(mj)}(\lambda) t_m
}R_{0}(\lambda,q)
e^{
\sum\limits_{j=1}^{\tilde N}\sum\limits_{m=1}^2 T^{(j)}  \tilde \rho^{(mj)}(\lambda) t_m
}
\end{eqnarray}
where  $R_0$ is  arbitrary $Q\times Q$ matrix function, $q=(q_1,\dots,q_N)$, parameters $q_i$ are complex in general and ${\cal{D}}$ is some integration region in space of  vector parameter $q$.

$N$-wave and Pohlmeyer equations correspond to 
$\rho^{(mj)}=0$, $\tilde \rho^{(mj)}(\lambda;x)= \lambda\delta_{mj}$ and  $\tilde \rho^{(mj)}=0$, $ \rho^{(mj)}(\lambda;x)= \lambda\delta_{mj}$ respectively.

The evident manifold of explicit particular solutions corresponds to the following  choice  of $R_0$:
\begin{eqnarray}\nonumber
(R_0(\lambda;q))_{\alpha\beta} = \left\{ \begin{array}{ll}\sum\limits_j  c^{(j)}_{\alpha\beta}(q)  \delta(\lambda-a^{(j)}_{\alpha\beta}), & \alpha\neq\beta
\cr
0,& \alpha=\beta  \end{array}\right., \;\;\alpha,\beta=1,\dots,Q
 \end{eqnarray}
 with scalar parameters $a^{(i)}_{\alpha_1\alpha_2}\neq 
 a^{(j)}_{\alpha_3\alpha_4}$
  $\forall \; i,j$ and  $(\alpha_1,\alpha_2)\neq   (\alpha_3,\alpha_4)$ and arbitrary functions 
 $c^{(j)}_{\alpha\beta}(q)$. Such solutions have the form of fractional rational function of the exponential functions  with linear in  $x$ arguments. 
   We do not consider these solutions in details. Instead, we concentrate on another manifold of particular solutions in the next subsection.    

\subsubsection{Dressing algorithm: reduction (\ref{red22}) and appropriate solutions to PDE0($t_m;n_0,w$), $m=1,2$, and PDE1($t_1,t_2;w^{(0)}$)}
 \label{sol_PDE0}
 \paragraph{Non-evolutionary part of the dressing algorithm.}
We represent the  non-evolutionary part of the dressing algorithm  describing the solutions manifold  to both  PDE0($t_m;n_0,w$), $m=1,2$, and PDE1($t_1,t_2;w^{(0)}$) assotiated  with reduction (\ref{red22}).
Solution of (\ref{red22}a)   should be taken 
 in the next form:
\begin{eqnarray}\label{H_1}\label{solH_1}
H_1(\mu;x)=\sum_{j=1}^{n_0} \hat h^{(j)}(x) \delta( \mu-b_{j}) ,
\end{eqnarray}
where $\hat h^{(i)}(x)$ are diagonal matrix functions and $b_{j}$, $j=1,\dots,n_0$ are scalar constants solving
the next linear algebraic system: 
\begin{eqnarray}\label{redH1_M_b}
b_{j}^{n_0}  = \sum_{i=1}^{n_0-1}b_{j}^{i}  r^{(i)},
\end{eqnarray}
where the parameters  $r^{(i)}$ are scalars. Requirement to have scalar $b_i$ and, 
as a consequence, scalar $r^{(i)}$ is related with transformation from the eq.(\ref{inter_0}) to the eq.(\ref{sol_psi-g-__}), see text after the eq.(\ref{R_def}).
Eq.(\ref{redH1_M_b}) has $n_0$ different roots $b_{j}$, $j=1,\dots,n_0$, which agrees with the summation limits in eq.(\ref{H_1}). 


We describe the manifold of the particular solutions to PDE1($t_1,t_2;w^{(0)}$) corresponding to 
the reduction (\ref{red22}) starting with the  original
system 
(\ref{u1},\ref{condition},\ref{U_sp30}a) modified by the reduction (\ref{red22}) as follows. 

Applying   $*\hat A^j*H_1$ to (\ref{u1}), $j=0,\dots,n_0-1$, one gets:
\begin{eqnarray}\label{arb_sol_sp4}
\Psi*{\bf V}=0,
\end{eqnarray} 
where ${\bf V}$ is $2Q\times n_0 Q$ matrix function
   (\ref{V_block}).
Similarly, applying $*\hat A^j*H_1$
to the eq.(\ref{condition}) one gets
\begin{eqnarray}\label{arb_con}
G(\lambda,\nu;x)*{\bf V}(\nu;x) = \tilde P(\lambda;x), 
\end{eqnarray}
where $\tilde P$ is a row of  $n_0$ blocks 
\begin{eqnarray}
&&
\tilde P =[H_1\;  \; \;
\hat A*H_1\;\cdots\; \hat A^{n_0-1}*H_1].
\end{eqnarray}
Finally, the spectral equation (\ref{U_sp30}a) 
can be written in 
the next form after applying  
$*\hat A^j*H_1$:
\begin{eqnarray}
A(\lambda,\nu)*V^{(j)}(\nu;x) &=&  V^{(j+1)}(\lambda;x)+  V^{(0)}(\lambda;x)
w^{(0j)},
\end{eqnarray} 
or
\begin{eqnarray}\label{VV}
A(\lambda,\mu)*{\bf V}(\mu;x)={\bf V}(\lambda;x) w(x),
\end{eqnarray}  
where $w$ is $n_0 Q\times n_0 Q$ block matrix (\ref{wwww}a).

In order to solve the system (\ref{arb_sol_sp4},\ref{arb_con},\ref{VV}) we 
assume diagonalizability of $w$, i.e. $w$ is representable in the form (\ref{PEP}).  The case when $w$ is not diagonalizable but 
reducible to Jordan form is more complicated ind will be
 discussed in different paper. 
 Now we describe all steps of the non-evolutionary part of the dressing  algorithm  using
 ${\cal{A}}$, $A$ and $\hat A$ given by the  eqs.(\ref{AAA0}).
 
 (\ref{1a}) coincides with the previous case, see eqs.(\ref{arb_sol_sp2},\ref{hatPsi}).
 
 (\ref{1b})
 Eq. (\ref{VV}) after 
multiplication by  $P$ from the right yelds the eq.(\ref{V_expl}), which we write in the next form: 
\begin{eqnarray}
\label{hatVV}
&&
\sum_{\gamma=1}^{n_0 Q}
{\bf V}_{\alpha\gamma}(\mu;x)P_{\gamma\beta}(x)=
\hat {\bf V}_{\alpha\beta}(x)\delta(\mu -E_\beta),\\\nonumber
&& \alpha=1,\dots,2 Q,\;\;\beta=1,\dots, n_0 Q,
\end{eqnarray}
where $\hat{\bf V}$ has the structure given by the eqs.(\ref{V_block2}) and
\begin{eqnarray}
\hat V^{(i)}(x)=\left[\begin{array}{c}\hat v^{(i)}_0(x) \cr
\hat v^{(i)}_1(x)\end{array}\right],\;\;i=0,\dots,n_0-1
.
\end{eqnarray}

(\ref{1c}) coincides with the previous case, see eqs.(\ref{GHH},\ref{sol_G}).

(\ref{1d}) 
Multiplying the   eq.(\ref{arb_sol_sp4}) by $P$ from the right, using the eq.(\ref{hatPsi}) 
for $\Psi$ and the 
eq.(\ref{hatVV}) for ${\bf V}$ one gets:
\begin{eqnarray}\label{arb_sol_sp42}
&&
 \sum_{\gamma=1}^{2Q}
{\hat\Psi}_{\alpha\gamma}(E_\beta;x)
{\hat {\bf V}}_{\gamma\beta}(x) =\sum_{j=0}^1\sum_{\gamma=1}^{2Q}({\hat\psi_j})_{\alpha\gamma}(E_\beta;x)
{(\hat {\bf V}_j)}_{\gamma\beta}(x)=0,\\\nonumber
&& \alpha=1,\dots, Q,\;\;\beta=1,\dots, n_0 Q,
\end{eqnarray} 
where
\begin{eqnarray}
\hat {\bf V}_{j} = 
[\hat v^{(0)}_j \;\;\cdots \;\; \hat v^{(n_0-1)}_j],
\;\;\;j=0,1.
\end{eqnarray}
Eq.(\ref{arb_con}) reads:
\begin{eqnarray}\label{arb_con_F}
 G(\lambda,\nu;x)* {\bf V}(\nu;x) =  H_1(\lambda;x)\hat P(\lambda),\;\;\;\hat P(\lambda) =
 [I\;\;\;\lambda I\;\cdots \;\lambda^{n_0-1}I] .
\end{eqnarray}
Multiplying this equation by $P$ from the right, substituting  $ G$ from the eq.(\ref{sol_G})
and
$ {\bf V}$ from the eq.(\ref{hatVV}) one gets: 
\begin{eqnarray}\label{arb_sol_sp5}
&&({ H_1})_\alpha(\lambda;x)\left(-\sum_{\gamma=1}^{2Q} 
\frac{({ H_{2}})_{\alpha\gamma}(E_\beta;x) 
{{{\hat {\bf V}}}}_{\gamma\beta}(x) }{\lambda-E_\beta}
-
\sum_{\gamma=1}^{n_0 Q}\hat P_{\alpha\gamma}(\lambda) P_{\gamma\beta}(x) \right)+\\\nonumber
&&
 \sum_{j=0}^{1}g_{0j}(\lambda;x) (\hat {\bf V}_{j}(x))_{\alpha\beta}\delta\left(\lambda-E_\beta)\right) 
=0,\;\;\;
\alpha=1,\dots, Q,\;\;\;\beta=1,\dots,n_0 Q.
\end{eqnarray}
Eq.(\ref{arb_sol_sp5})  consists of  terms with separated spectral parameters. Thus,
it must be splitted into two equations:
\begin{eqnarray}\label{arb_sol_sp51}\label{arb_sol_sp5_ex1_0}
&&({ H_1})_\alpha(\lambda;x)\left(\sum_{\gamma=1}^{2Q} 
\frac{{( H_{2})}_{\alpha\gamma}(E_\beta,;x) 
\hat {\bf V}_{\gamma\beta}(x)}{\lambda-E_\beta} 
+
\sum_{\gamma=1}^{n_0 Q}\hat P_{\alpha\gamma}(\lambda) P_{\gamma\beta}(x) \right)=0,
\\\label{arb_sol_sp52}\label{s2_ex1_0}
&&
 \sum_{j=0}^{1}g_{0j}(E_\beta;x) (\hat {\bf V}_{j}(x))_{\alpha\beta}
=0,\;\;
\alpha=1,\dots, Q,\;\;\;\beta=1,\dots,n_0 Q.
\end{eqnarray}
Consider $H_1$ defined by the eq.(\ref{solH_1}). 
Then the equation (\ref{arb_sol_sp51}) becomes  equivalent to the next one:
\begin{eqnarray}\label{arb_sol_sp5_ex1}
&&\sum_{j=0}^1\sum_{\gamma=1}^{2Q} 
{( h_{2j})}_{\alpha\gamma}(E_\beta;x)  \left(E_\beta -b_j\right)^{-1}
(\hat {\bf V}_j)_{\gamma\beta}(x)
-
\sum_{\gamma=1}^{n_0 Q}\hat P_{\alpha\gamma}( b_j) P_{\gamma\beta}(x) =0.
\end{eqnarray}
Eqs. (\ref{arb_sol_sp42}, \ref{arb_sol_sp5_ex1}) must be viewed as a system of equations for $\hat {\bf V}$, while eq. (\ref{s2_ex1_0}) 
is a matrix equation for  elements of  $E$ and  $P$.
These three equations may be written in terms of $w$ 
in the following way.
First of all, the  eq.(\ref{arb_sol_sp42})   relates ${\hat{\bf V}}_j$, $j=0,1$: ${\hat{\bf V}}_1(x)=-\hat \psi_{1}^{-1}(E_\beta;x) \hat \psi_{0}(E_\beta;x) {\hat{\bf V}}_0(x)$. 
Substituting this relation into 
 eq.(\ref{s2_ex1_0}) and into eq.(\ref{arb_sol_sp5_ex1}) multiplied by $(E_\beta-b_j)$ from the right one results in
\begin{eqnarray}\label{sol_sp422__}
&&
\sum_{\gamma=1}^{2Q}
({\psi^{(-)}})_{\alpha\gamma}(E_\beta;x)
({\hat {\bf V}}_0(x))_{\gamma\beta}=0,\\\nonumber
&&
\alpha=1,\dots, Q,\;\;\;\beta=1,\dots,n_0 Q.
\\\label{sol_sp423__}
&&
\sum_{\gamma=1}^{Q}h^{(-)}_{\alpha\gamma}(E_\beta;x)
 ({\hat{\bf V}}_{0})_{\gamma\beta}(x) 
= \sum_{\gamma=1}^{n_0 Q} \hat P_{\alpha\gamma}(b_j) P_{\gamma\beta}(x)  (E_\beta-b_j ),
\\\nonumber
&&
\alpha=1,\dots Q,\;\;\beta=1,\dots,n_0Q,\;\;j=1,\dots,n_0,
\end{eqnarray}
where
\begin{eqnarray}\label{psi_h__}
&&
\psi^{(-)}(\lambda;x)={ g_{00}}(\lambda;x)
 - 
{g_{01}}(\lambda;x) \hat \psi_{1}^{-1}(\lambda;x) 
\hat \psi_{0}(\lambda;x)
,\\\nonumber
&&
h^{(-)}(\lambda;x) ={
 h_{20}}(\lambda;x)   - 
 { h_{21}}(\lambda;x)\hat\psi_{1}^{-1}(\lambda;x) \hat\psi_{0}(\lambda;x).
\end{eqnarray}
Next, eliminating $\hat {\bf V}_{0}$ from the eq. (\ref{sol_sp422__}) using the eq.(\ref{sol_sp423__}) one gets
\begin{eqnarray}\label{inter_0}
&&
\sum_{\gamma_1=1}^{Q}\sum_{\gamma_2=1}^{n_0 Q} R_{\alpha\gamma_1}(E_\beta;x) E_\beta^{-1}\hat P_{\gamma_1\gamma_2}(b_j) P_{\gamma_2\beta}(x)  (E_\beta-b_j)  =0,
\\\nonumber
&&
\alpha=1,\dots Q,\;\;\beta=1,\dots,n_0Q.
\end{eqnarray}
where
\begin{eqnarray}\label{R_def}
R(\lambda;x)=\psi^{(-)}(\lambda;x) (h^{(-)})^{-1}(\lambda;x)\lambda.
\end{eqnarray}
Finally, applying $P^{-1}$ from the right,  using the fact that $b_j$ are scalars and  the evident identity
\begin{eqnarray}\label{b_identity}
\hat P(b_j)\;(w-b_j I) \equiv \hat P(0) \;w
\end{eqnarray}
one gets the resulting equation  in terms of $w$: 
\begin{eqnarray}\label{sol_psi-g-__}
 \Big[\sum_{\gamma=1}^{Q}
R_{\alpha\gamma}(w;x) \Big]_{\gamma\beta} =0,
\;\;\;\alpha=1,\dots Q,\;\;\;\beta=1,\dots,n_0Q,
\end{eqnarray}
which  is an equation for the  elements of    $Q\times Q$ matrices $w^{(0j)}$, $j=0,\dots,n_0-1$. This ends the non-evolutionary part of the dressing algorithm. 

We see that, along with the solution $U(\lambda,\mu;x)$ to the spectral problem, we have described implicitly the function $w(x)$ (solution to the apporpriate PDE0) and, as a consequence, the function $w^{(0)}(x)$ (solution to PDE1).
As a simple example, 
let $n_0=1$. Then  $\hat A*H_1=0$, $\tilde P(k)=I$, $w=w^{(00)}$. The eq. (\ref{sol_psi-g-__}) gets the next form: 
\begin{eqnarray}\label{sol_psi-g-}
&& \Big[\sum_{\gamma=1}^{Q}R_{\alpha\gamma}(w;x) \Big]_{\gamma\beta} =0
,\;\;\alpha,\beta=1,\dots,Q.
\end{eqnarray}

Eq.(\ref{sol_psi-g-__}) holds for any PDE1 and assotiated PDE0.
In order to show the equivalence of this equation to the non-differential equations derived in Sec.\ref{Algebraic} (see eqs.(\ref{w_C},\ref{w_C_G_P},\ref{w_C_G_Pred})),
we must introduce the particular dependence of $R$  on variables $t_i$ and $x_i$, which will be done in the rest of Sec.\ref{Evolution}.

\paragraph{Evolutionary part of the dressing algorithm.}
\label{Evolution}
As we have seen,  the structure of eq.(\ref{sol_psi-g-__})
 does not depend 
on the particular $x$-dependence of the functions $\psi^{(-)}$ and $h^{(-)}$ defined through the $x$-dependence of the  functions  $\Psi$, $G_0$ and $H_2$ (see eqs.(\ref{x},\ref{G_x},\ref{case01},\ref{case02})).
 Here we describe the $x$-dependence of (\ref{sol_psi-g-__}), related with the $x$-dependence of $\Psi$ (eq.(\ref{x})), $G_0$ (eq.(\ref{G_x}))  and $H_2$ (eqs.(\ref{H22}) or (\ref{H22_2})). Remark, that evolution of $H_1$ given by the eqs.(\ref{H12}) or (\ref{H12_2}), is not needed in resulting formula (\ref{sol_psi-g-__}).  
 Emphasise that index $m$ appearing in the equations of this section is meaningful only in the case (\ref{case02}), when 
${\cal{P}}^{(m)}=\hat P^{(m)}=0$ and $P^{(m)}=0$. Otherwise this index must be fixed indicating that the appropriate PDEs do not possess the commuting flows.

Functions ${\cal{C}}^{(m)}$, $C^{(m)}$ and $\hat C^{(m)}$ are defined by the eqs.(\ref{PsiP2},\ref{PsiP3},\ref{case01b}):
\begin{eqnarray}\label{CCC}
{\cal{C}}^{(m)} (\lambda,\mu)&=&\hat C^{(m)}(\lambda,\mu)=
-\Big(\rho(\lambda)\delta(\lambda-\mu)\Big)_\lambda I,
\\\nonumber
C^{(m)} (\lambda,\mu)&=&
-\Big(\rho(\lambda)\delta(\lambda-\mu)\Big)_\lambda
I_{2}.
\end{eqnarray}
Eqs.(\ref{x}) (in view of  eq.(\ref{hatPsi})) and (\ref{G_x}) yield the following equations for the functions $\hat \psi_j$ and  $\tilde g_{0j}$  respectively ($j=0,1$):
\begin{eqnarray}\label{psi_diff}
&&\Big({\hat \psi_i}(\lambda;x)\Big)_{t_m} +\sum_{j=1}^N \rho^{(mj)}(\lambda)
 \Big({\hat \psi_i} (\lambda;x) \Big)_{x_j}
-\Big(\hat \psi_i(\lambda;x)\rho^{(m0)}
(\lambda)\Big)_{\lambda}=0,\;\;i=0,1,
\end{eqnarray}
\begin{eqnarray}\label{g0_diff}
&&\Big({ g_{0i}}(\lambda;x)\Big)_{t_m} +\sum_{j=1}^N \rho^{(mj)}(\lambda)
\Big( { g_{0i}} (\lambda;x)\Big)_{x_j}
-\Big(g_{0i}(\lambda;x)\rho^{(m0)}
(\lambda)\Big)_{\lambda} +\\\nonumber
&&
 \sum_{j=1}^{\tilde N}T^{(j)} \tilde \rho^{(mj)}(\lambda)\; { { g_{0i}}}(\lambda;x) =0,\;\;i=0,1,
\end{eqnarray}
Combining these equations and taking into account the definition of $\psi^{(-)}$ ( eq.(\ref{psi_h__}a)) we obtain equation for $\psi^{(-)}$:
\begin{eqnarray}\label{psi__diff}
&& \psi^{(-)}_{t_m}(\lambda;x) +\sum_{j=1}^N \rho^{(mj)}(\lambda)
\psi^{(-)}_{x_j} (\lambda;x)
- \Big(\psi^{(-)} (\lambda;x)\rho^{(m0)}
(\lambda)\Big)_\lambda+ \\\nonumber
 &&
 \sum_{j=1}^{\tilde N}T^{(j)} \tilde \rho^{(mj)}(\lambda)\; 
  \psi^{(-)}(\lambda;x)=0,\;\;i=0,1.
\end{eqnarray}

Mostly general equation for $H_2$ is eq.(\ref{H22}), which may be written in the next form in view of eq.(\ref{case01}b):
\begin{eqnarray}
{H_2}_{t_m}+\sum_{j=1}^N{H_2}_{x_j} *A^{(mj)}+\sum_{j=1}^{\tilde N} T^{(j)} H_2 *\tilde A^{(mj)}- H_2 *A^{-1}*C^{(m)}*A=0.
\end{eqnarray}
Then, in particular, eq.(\ref{H22_2}) corresponds to $C^{(m)}=0$. 
Introduce the function ${\cal{H}}(\lambda;x)$ by the formula:
\begin{eqnarray}\label{calH}
H_2={\cal{H}}_2*A.
\end{eqnarray}
Then the above equation takes the next form after applying $*A^{-1}$:
\begin{eqnarray}\label{H_2_M}
{{\cal{{\cal{H}}}}_2}_{t_m}+\sum_{j=1}^N{{\cal{H}}_2}_{x_j} *A^{(mj)} +\sum_{j=1}^{\tilde N}T^{(j)} {\cal{H}}_2 *\tilde A^{(mj)}- {\cal{H}}_2 *C^{(m)}=0.
\end{eqnarray}
Function ${\cal{H}}_2$ has the structure analogous to $H_2$:
\begin{eqnarray}
{\cal{H}}_2(\mu;x)=[{\rm h}_{20}(\mu;x) \;\;\;{\rm h}_{21}(\mu;x)].
\end{eqnarray}
Then
eqs.(\ref{H_2_M}) 
yields:
\begin{eqnarray}\label{G_diff}
&&
\Big({ {\rm h}_{2i}}(\lambda;x)\Big)_{t_m} +
\sum_{j=1}^N \rho^{(mj)}(\lambda)\; \Big({{\rm h}_{2i}}(\lambda;x)\Big)_{x_j} +\\\nonumber
&&
\sum_{j=1}^{\tilde N}T^{(j)} \tilde \rho^{(mj)}(\lambda)\; { {\rm h}_{2i}}(\lambda;x)-
\Big( {\rm h}_{2i}(\lambda;x)\Big)_{\lambda}
 \rho^{(m0)}(\lambda)
=0,\;\;i=0,1.
\end{eqnarray}
Combining eqs.(\ref{psi_diff}) and (\ref{G_diff}), taking into 
account definition of $h^{(-)}$ (\ref{psi_h__}b) one gets
\begin{eqnarray}\label{h__diff}
&&
{ {\rm h}^{(-)}_{t_m}}(\lambda;x) +
\sum_{j=1}^N \rho^{(mj)}(\lambda)\; {{\rm h}^{(-)}}_{x_j}(\lambda;x) +\\\nonumber
&&
\sum_{j=1}^{\tilde N}T^{(j)} \tilde \rho^{(mj)}(\lambda)\; { {\rm h}^{(-)}}(\lambda;x)-
 \Big({\rm h}^{(-)}(\lambda;x)\Big)_\lambda
 \rho^{(m0)}(\lambda)
=0,\;\;i=0,1.
\end{eqnarray}
where 
\begin{eqnarray}
{\rm h}^{(-)}(\lambda;x)= {\rm h}_{20}(\lambda;x)-{\rm h}_{21}(\lambda;x) \hat\psi_1^{-1}(\lambda;x)\hat\psi_0(\lambda;x),\;\;\;
h^{(-)}(\lambda;x) =
 {\rm h}^{(-)}(\lambda;x) \lambda .
 \end{eqnarray}
 Finally, linear PDE for $R(\lambda;x) $ defined by the eq.(\ref{R_def})  
 reads:
 \begin{eqnarray}\label{R__diff}
&&
{ R_{t_m}}(\lambda;x) +
\sum_{j=1}^N \rho^{(mj)}(\lambda)\; R_{x_j}(\lambda;x) +\\\nonumber
&&
\sum_{j=1}^{\tilde N} \tilde \rho^{(mj)}(\lambda)\; [T^{(j)},R(\lambda;x)]-
 \Big(R(\lambda;x)
 \rho^{(m0)}(\lambda)\Big)_\lambda
=0,\;\;i=0,1.
\end{eqnarray}
 Thus, evolutionary part of the dressing algorithm, nn.\ref{2a},\ref{2b}, is reduced to the solution of the single eq.(\ref{R__diff}).
We describe  two particular examples.

\paragraph{Solutions to the PDE0s (\ref{rho_w}).}
\label{Sol:Case03}
In this case $n_0=1$ so that $w\equiv w^{(0)}$. 
It has been shown in Sec.\ref{Section:matr}, that
 index $m$ must be fixed. Let $m=1$. Use the following notations: 
 \begin{eqnarray}\label{rho_simple}
 t\equiv t_1, \;\;\rho^{(j)}\equiv \rho^{(1j)},  \;\;\tilde \rho^{(j)}\equiv \tilde \rho^{(1j)}.
 \end{eqnarray}
In general, when $\rho^{(0)}\neq 0$, 
the eq.(\ref{R__diff}) may be integrated by the method of characteristics. Eq.(\ref{R__diff}) yelds:
\begin{eqnarray}\label{char1}
&&
\frac{d\lambda}{dt}=-\rho^{(0)}(\lambda)
,\\
\nonumber
&&
\frac{dx_j}{dt}=\rho^{(j)}(\lambda),\;\;j=1,\dots,N,
\\
\nonumber
&&
\frac{dR}{dt}=\rho^{(0)}_\lambda(\lambda) R-
\sum_{j=1}^{\tilde N} \tilde \rho^{(j)}(\lambda)\; 
  [T^{(j)},R].
\end{eqnarray}
Integrating the eqs.(\ref{char1}) in quadratures we receive:
\begin{eqnarray}\label{char_psi}
&&
t=-\int\limits^\lambda \frac{d\lambda}{\rho^{(0)}(\lambda)}+c_{0},\\
\nonumber
&&
x_j=\int\limits^t \frac{dt}{\rho^{(0)}(\lambda(t))}+c_{j},\;\;j=1,\dots,N,\\
\nonumber
&&
R=e^{-\int\limits^{t} d t
 \sum\limits_{j=1}^{\tilde N}T^{(j)}\tilde\rho^{(j)}(\lambda(t))}
 \frac{C(c_{0},\dots,c_{N})}{\rho^{(0)}(\lambda)}
 e^{\int\limits^{t} d t
 \sum\limits_{j=1}^{\tilde N}T^{(j)}\tilde\rho^{(j)}(\lambda(t))} ,
\end{eqnarray}
where $c_{i}$ are integration constants and $C(z_0,\dots,z_N)$ is an arbitrary function of $N+1$  arguments.

Integration constants $c_i=c_i(\varkappa,x)$, $i=0,\dots,N$,  must be expressed  in terms of $\lambda$ and $x$ using (\ref{char_psi}a,b).
Let 
\begin{eqnarray}
C(z_0,\dots,z_N) = z_0 - F(z_0,\dots,z_N).
\end{eqnarray}
Then the eq.(\ref{sol_psi-g-__}) yields
\begin{eqnarray}\label{sol:int_0}
&& ({c_{0}(w,x)})_{\alpha\beta} =\\\nonumber
&&
\sum_{\gamma=1}^{Q} \Big[
e^{-\left.\left(\int\limits^{t} d t
 \sum\limits_{j=1}^{\tilde N}T^{(j)}_\alpha\tilde\rho^{(j)}(\lambda(t))\right)\right|_{\lambda\to w}}
 F_{\alpha\gamma}(c_{0}(w;x),\dots,c_{N}(w;x))  
e^{\left.\left(\int\limits^{t} d t
 \sum\limits_{j=1}^{\tilde N}T^{(j)}_\gamma\tilde\rho^{(j)}(\lambda(t))\right)\right|_{\lambda\to w}} \Big]_{\gamma\beta} =0,
\\\nonumber
&&
\alpha=1,\dots Q,\;\;\beta=1,\dots,n_0Q.
\end{eqnarray}
This equation describes the solutions space to  the PDE0
(\ref{rho_w}) with  $m=1$ and notations (\ref{rho_simple}). 
Of course, the same result may be obtained using the algebraic approach.

\paragraph{Construction of the solutions manifold  to the PDE1($t_1,t_2;w^{(0)}$) and PDE0($t_m;n_0,w$), $m=1,2$, derived in  Sec.\ref{Char_Gen}.
}
\label{Char_Sol}
If $\rho^{(m0)}=0$ (see eq.(\ref{case02})) then 
eq.(\ref{R__diff})  may be  solved explicitly:
\begin{eqnarray}\label{psi_h}
R(\lambda;x)=\int\limits_{{\cal{D}}} dq\;e^{I\sum\limits_{j=1}^N \Big(q_j x_j - \sum\limits_{m=1}^2 \rho^{(mj)}(\lambda) q_j t_m\Big)- 
\sum\limits_{j=1}^{\tilde N}\sum\limits_{m=1}^2 T^{(j)}  \tilde \rho^{(mj)}(\lambda) t_m}R_0(\lambda,q)e^{\sum\limits_{j=1}^{\tilde N}\sum\limits_{m=1}^2 T^{(j)}  \tilde \rho^{(mj)}(\lambda) t_m},
 \end{eqnarray}
 where $q=(q_1,\dots,q_N)$, parameters $q_i$ are complex in general and integration is over some region of the space of  parameter $q$. 
The proper choice of $R_0$ is following:
\begin{eqnarray}
R_0(\lambda,q)=\lambda-\tilde F(\lambda,q).
\end{eqnarray}
Then the eq.(\ref{sol_psi-g-__}) gives us 
\begin{eqnarray}\label{w_C_G}
&&
w_{\alpha\beta}= \sum_{\gamma=1}^Q \Big[
e^{ - 
\sum\limits_{m=1}^2\sum\limits_{j=1}^{\tilde N} T^{(j)}_\alpha  \tilde \rho^{(mj)}(w) t_m
}
 F_{\alpha\gamma}\Big( w,x_1 I_{n_0} - \sum\limits_{m=1}^2 \rho^{(m1)}(w)  t_m ,\dots,\\\nonumber
&&
x_N I_{n_0}- \sum\limits_{m=1}^2 \rho^{(mN)}(w)  t_m \Big) 
e^{  
\sum\limits_{m=1}^2\sum\limits_{j=1}^{\tilde N} T^{(j)}_\gamma  \tilde \rho^{(mj)}(w) t_m
} \Big]_{\gamma\beta}
,
\\\nonumber
&&
\alpha=1,\dots,Q,\;\;\beta=1,\dots, n_0 Q,
\end{eqnarray}
where
 we use notation
\begin{eqnarray}
F(z_0,\dots,z_N)=\int d q \tilde F(z_0,q_1,\dots,q_N) e^{i \sum\limits_{j=1}^Nq_j z_j }.
\end{eqnarray}

\paragraph{ Solutions to $N$-wave equation.}
 
If $\rho^{(mi)}=0$, $\tilde \rho^{(mj)}=A\delta_{mj}$,  then PDEs have derivatives only with respect to $t_i$ (but not  with respect to $x_i$).
We may take
$F(z_0,z_1,\dots,z_N)=F(z_0)$ in the eq.(\ref{w_C_G}), so that 
 eq.(\ref{w_C_G}) gives us the algebraic equation (\ref{w_C}).

\paragraph{ Solutions to Pohlmeyer equation.}
 One must substitute 
$ \rho^{(mj)}= A\delta_{mj}, \;\; T^{(i)}=0,\;\; N=2
$
  into the eq.(\ref{w_C_G}), which yelds the  algebraic equation (\ref{w_C_G_P}) describing solution manifold to the eq.(\ref{P}).
Solution manifold to the eq.(\ref{P_red}a) corresponds to the  reduction
\begin{eqnarray}
\tilde F(z_0,q_1,q_2) = \tilde F(z_0,q_1)\delta(q_2-z_0 q_1),\;\;x_2=0,
\end{eqnarray}
which  yields the algebraic equation (\ref{w_C_G_Pred}).

\subsection{Second version of the dressing method for PDE1($t_1,t_2;w^{(0)}$) with reduction (\ref{red2},\ref{red2_r}).}
\label{Second}
Disadvantage of the dressing algorithm described above is a poor available solutions space to PDE0s, since all freedom is assotiated  with the function  $R$ which is $Q\times Q$ matrix function, while $w$ has $n_0Q^2$ scalar fields.  However, we know that the  algebraic approach provides a fullness of the solutions space. Thus we may ask if there is another version of the dressing method which yelds fullness of the solutions space as well.

Such version of the dressing method is based on the integral equation 
(\ref{u1}) as $n_0 Q\times n_0Q$ matrix equation (it was $Q\times Q$ matrix equation in the above algorithm). For simplicity, we consider only those PDE0s which 
are assotiated with PDE1s. Thus, PDE0 (\ref{rho_w}) with nonzero $\rho^{(m0)}$ is not the 
subject of this subsection.

This version of the dressing  algorithm is very similar to one discussed
in Secs.\ref{Section:M_1},\ref{Solutions}. In particular, all equations for dressing and spectral functions remain correct, only matrix dimensionalities must be changed.   So we just underline a few features. 

\begin{enumerate}
\item It follows  from the first paragraph of this section, that  PDE0($t_m;n_0,w$), $m=1,2$, with any fixed $n_0$ is assotiated with its own integral equation (\ref{u1}). 
\item Field 
\begin{eqnarray}\label{w_2}
w = H_2*U*H_1
\end{eqnarray}
must have   the structure 
(\ref{wwww}a).  As before, $w^{(0)}$ is a solution of PDE1. However, $r^{(i)}$ are not necessary scalars unlike Secs.\ref{Section:M_1} and \ref{Solutions}. Since structure (\ref{wwww}a) must be supported by the nonlinear  PDEs (\ref{equ-w_com}), 
 parameters  $r^{(i)}$ should be  defined as in the algebraic approach, see a paragraph after the eqs.(\ref{wwww}).  
\item\label{32} In order to establish  relation between PDE1($t_1,t_2;w^{(0)}$) and 
PDE0($t_m;n_0,w$), $m=1,2$, we take the  reduction  (\ref{red22}) in the next  form:
\begin{eqnarray}\label{red2_2}
\hat A*H_1=0 \;\;\;\Rightarrow \;\;\; H_2*U*\hat A*H_1 =0.
\end{eqnarray}
Now the functions ${\cal{A}}$, $A$ and $\hat A$ are defined as follows (compare with the formulae (\ref{AAA0})):
\begin{eqnarray}\label{AAA2}
{\cal{A}}(\lambda,\mu) =\hat A(\lambda,\mu)= \lambda
\delta(\lambda-\mu)I_{n_0}
,\;\;\;\;
A(\lambda,\mu)=
\lambda\delta(\lambda-\mu)I_{2n_0}.
\end{eqnarray}
\item\label{322} Similarly to the n.\ref{32},  in order to establish  relation between PDE1($t_1,t_2;w^{(0)}$) and 
PDE0($t_m;k_0,\tilde w$), $m=1,2$, we take the  reduction  (\ref{red11}) in the next  form:
\begin{eqnarray}\label{red1_2}
H_2*A=0 \;\;\;\Rightarrow \;\;\; H_2*A*U*H_1 =0.
\end{eqnarray}
\item\label{34} Owing to n.\ref{32}, derivation of PDE0($t_m;n_0,w$) in this section is equivalent to  derivation 
made in  Sec.\ref{Section:matr} with $\rho^{(m0)}=0$, see resulting equation (\ref{rho_w}). This 
fact suggests us the general form of PDE0($t_m;n_0,w$), eq.(\ref{wdir}). 
\item\label{342} Similarly, owing to n.\ref{322}, we may derive general form of  PDE0($t_m;k_0,\tilde w$), see eq.(\ref{twdual}). We do not represent appropriate calculations.
\item\label{35}
Eq.(\ref{w_C_G}) 
must be replaced by the next equation
\begin{eqnarray}\label{w_C_G_2}
&&
w_{\alpha\beta}= \sum_{\gamma=1}^{n_0 Q} \Big[ e^{ - 
\sum\limits_{m=1}^2\sum\limits_{j=1}^{\tilde N} T^{(j)}_\alpha  \tilde \rho^{(mj)}(w) t_m}
F_{\alpha\gamma}\Big( w,x_1 I_{n_0} - \sum\limits_{m=1}^2 \rho^{(m1)}(w)  t_m ,\dots,\\\nonumber
&&
x_N I_{n_0}- \sum\limits_{m=1}^2 \rho^{(mN)}(w)  t_m \Big) e^{  
\sum\limits_{m=1}^2\sum\limits_{j=1}^{\tilde N} T^{(j)}_\gamma  \tilde \rho^{(mj)}(w) t_m}\Big]_{\gamma\beta},
\\\nonumber
&&
\alpha=1,\dots,Q,\;\;\beta=1,\dots, n_0 Q.
\end{eqnarray}
\end{enumerate}
We see that the algorithm discussed in this section provides the same richness of the solutions space as the algebraic approach does. Namely, we have $n_0 Q^2$ arbitrary functions $F_{\alpha\beta}$, $\alpha=1,\dots,Q$, $\beta=1,\dots n_0Q$, and 
diagonal ($T^{(j)}\neq 0$) or arbitrary ($T^{(j)}= 0$)) matrix  parameters  $r^{(j)}$, $j\ge 1$.

\section{Conclusions}
\label{Conclusions}

First of all,
we have represented the algorithm for the  derivation of the system of non-differential equations which  implicitly describes the special type of  solutions to Self-dual type  $S$-integrable equations (\ref{SP_s}).
This  is possible due to the relation  between $S$-integrable PDEs (PDE1s)
and appropriate family of matrix first order  quasilinear PDEs (PDE0s).  

As we have seen, PDE0s have set of  remarkable properties:
\begin{enumerate}
\item
Manifold of  PDE0s is splitted into  subclasses. Each subclass  joins  such PDE0s which are  
 lower dimensional reductions of appropriate  PDE1. 
 These PDE0s admit infinitely many commuting flows, similar to PDE1s. Besides, there is a subclass of PDE0s which is not related with  PDE1s; we have not identified the commuting flows for this subclass.
\item
Any PDE0 admits the spectral system consisting of two linear  equations for the spectral 
 function ${\bf V}(\lambda;x)$. But only 
one of these equations has derivatives with respect to variables 
$x_i$ and $t_i$, unlike the   spectral systems for PDE1s. This fact gives rise to  all 
following properties written in this list. 
\item
The solutions space of PDE0s  is implicitly described by the system of non-differential equations. This  is a typical description for the first order quasilinear equations \cite{ts2,ts1,dn,SZ1,ZS2}.
\item 
As we know, PDE1s (like any $S$-integrable PDE)  may be written using one of the following procedures: (a) applying $H_2 * A^n*$ and $*\hat A^j*H_1$ to the spectral problem and using the definitions of the dependent variables 
in terms of the spectral and dressing functions, 
see eq.(\ref{ww}),
 or (b)  as the compatibility conditions of the appropriate spectral problems. However, PDE0s may be derived only using the first procedure.
\item\label{xdep}
Functions $\rho^{(mj)}(z)$ and $\tilde \rho^{(mj)}(z)$ may be replaced by $\rho^{(mj)}(z;x)$ and  $\tilde \rho^{(mj)}(z;x)$ respectively, which are representable by positive power series of $z$ with  {\it scalar} coefficients explicitly  depending on $x$. The solutions spaces to such PDEs is implicitly described by the system of the  non-differential equations supplemented by the pair of  decoupled linear PDEs  with non-constant coefficients which, in turn, are integrable by the method of characteristics. We leave this statement without proof.
\end{enumerate}

Although all basic conclusions have been established  using the  algebraic approach, we represent a version of the dressing method allowing to derive the same system of non-differential equations describing solutions space to PDE0s and PDE1s. In addition, this method  reproduces the manifold of classical solutions to PDE1s.

 The fact that the dressing method exhibits a connection with the method of characteristics is important for unification of the methods for solving of nonlinear PDEs in multidimensions.
It tells us that the dressing method is more flexible in comparison, at least, with some other integration methods mentioned in this paper,
such as Inverse Spectral Transform (IST) ($S$-integrability),  direct linearization ($C$-integrability) and method of characteristics.   In fact, direct linearization may not be applied to both  $S$-integrable equations and PDE0s. Similarly, 
IST may not be applied to PDE0s and to $C$-integrable equations.
Method of characteristics  was modified to hodograph and generalized hodograph method allowing to study some ($S$-integrable) PDEs by the method of hydrodynamic reductions. However, using this technique we face  a problem of the explicit representation of  solutions. 

On the contrary,  although originally the dressing method was invented as a special technique for solving $S$-integrable equations, it was shown in \cite{ZS2} as well as in this paper that the properly modified  dressing method works also for PDE0s. It yelds  exactly the same matrix non-differential equation describing solutions space as method of characteristics does. The fact that the dressing method may be a starting point for solving $C$-integrable equations has been demonstrated in \cite{Z} as well as in \cite{ZS2}. 


Author thanks Professors P.M.Santini and A.B.Shabat for useful 
discussions and Professor I.M.Krichever for important comments. 
The work was supported by INTAS Young Scientists Fellowship Nr. 
04-83-2983, RFBR grants 04-01-00508, 06-01-90840, 06-01-92053 
and grant NS 7550.2006.2.

\end{document}